# Conditional Uncertainty Quantification of Stochastic Dynamical Structures Considering Measurement Conditions

Feng Wu *, Yuelin Zhao, Li Zhu

State Key Laboratory of Structural Analysis, Optimization and CAE Software for Industrial Equipment, School of Mechanic and Aerospace Engineering, Dalian University of Technology, Dalian 116024, China

*Corresponding author: wufeng_chn@163.com

**Abstract:** How to accurately quantify the uncertainty of stochastic dynamical responses affected by uncertain loads and structural parameters is an important issue in structural safety and reliability analysis. In this paper, the conditional uncertainty quantification analysis for the dynamical response of stochastic structures considering the measurement data with random error is studied in depth. A method for extracting the key measurement condition, which holds the most reference value for the uncertainty quantification of response, from the measurement data is proposed. Considering the key measurement condition and employing the principle of probability conservation and conditional probability theory, the quotient-form expressions for the conditional mean, conditional variance, and conditional probability density function of the stochastic structural dynamical response are derived and are referred to as the key conditional quotients (KCQ). A numerical method combining the non-equal weighted generalized Monte Carlo method, Dirac function smoothing technique, and online-offline coupled computational strategy is developed for calculating KCQs. Three linear/nonlinear stochastic dynamical examples are used to verify that the proposed KCQ method can efficiently and accurately quantify the uncertainty of the structural response considering measurement conditions. The examples also compare the traditional non-conditional uncertainty quantification results with the conditional uncertainty quantification results given by KCQs, indicating that considering measurement conditions can significantly reduce the uncertainty of the stochastic dynamical responses, providing a more refined statistical basis for structural safety and reliability analysis.

## 1. Introduction

The analysis of structural vibration response under time-varying loads has been an important task in the fields of mechanics, engineering, and beyond [1, 2]. Constructing efficient and high-precision time integration algorithms to predict the process of structural displacement, velocity, and other responses changing over time has always been one of the core research directions in the field of computational mechanics [3, 4]. In this direction, a variety of time integration algorithms have been developed, such as the Newmark method [5], the symplectic method [6, 7], the precise integration method [8-10], and so on. However, these methods mainly aim to solving deterministic structural dynamical response problems. In real engineering, structures are inevitably affected by various uncertain factors. For example, assembly errors during the construction process often leads to uncertainties in structural geometrical parameters [11, 12]; material preparation processes often lead to uncertainties in material parameters [13-15]; earthquakes [16], wind [17, 18], ocean waves [19, 20], etc., often lead to uncertainties in structural loads [21, 22]. For the assessment of the impact of these uncertain factors on structural responses, traditional deterministic algorithms are no longer applicable, and it is necessary to develop efficient methods for the uncertainty quantification of stochastic structural response [23]. In many engineering problems, engineers usually arrange measurement points on the structure to measure structural responses [24]. When quantifying the stochastic dynamical response of the structure with random parameters, if measurement data conditions can be considered, it is clear that the uncertainty quantification of the stochastic response can be further refined, providing a more reliable basis for structure safety and reliability analysis.

However, in actual engineering problems, measurement errors caused by human operation and instrument errors also have randomness [25]. Therefore, how to accurately and efficiently quantify the uncertainty of the dynamical response of stochastic linear/nonlinear structures given measurement conditions with random errors is an extremely important issue.

It is noteworthy that extensive and in-depth research has been conducted on structural dynamical response problems involving uncertain parameters. Existing work can be broadly categorized into two types: one focuses on how to accurately describe structural uncertainties [26, 27], while the other addresses how to construct algorithms for efficiently and accurately estimating the structural response uncertainty when the structural uncertainty model is known [10]. The first category aims to establish mathematical models describing structural uncertainty, with significant achievements such as the Karhunen-Loève series expansion for random fields [28], the maximum entropy method for the uncertainty parameter with limited samples [21, 29, 30], the interval model under incomplete cognitive conditions [31-33], and so on. For further reading on this topic, interested readers are referred to Refs. [34-36]. The second category aims to quantify the uncertainty of structural response. This has been a popular research area in the field of stochastic computational mechanics, yielding various effective uncertainty quantification methods, including the Monte Carlo (MC) method [37], the stochastic perturbation method [38, 39], the probability density evolution method [40-43], the stochastic perturbation collocation method [28, 44, 45], the generalized polynomial chaos expansion method [46], the surrogate model [47], and so on. Among these methods, the MC method is widely used due to its simplicity and ease of integration with deterministic structural analysis programs. Subsequent researchers have proposed the quasi-MC method, which achieves faster convergence than the traditional MC method by employing low-discrepancy sample points instead of random ones [48-50]. However, traditional quasi-MC method assign an equal integration weight, i.e., the reciprocal of sample size, for each sample. In Refs. [51, 52], Chen and Li et al. pointed out that assigning different weights for different samples based on the distribution characteristics of samples can further enhance the accuracy of the quasi-MC method. They introduced the concept of generalized discrepancy and utilized the Voronoi diagrams to generate the sample set and weights. Subsequently, Wu and Huang et al. [53] proposed a novel method for generating low-discrepancy sample sets with non-equal weights by combining the least squares method with the generalized polynomial chaos expansion method. Wu and Zhao et al. [54] introduced the generalized L2 discrepancy based on general points (GL2D-GP). Based on the principle of minimizing GL2D-GP and the permutation-based generalized Halton sequence, they employed the evolutionary algorithm to construct the generalized Halton sequence with non-equal weights. The above studies demonstrate that, the integration formats with non-equal weights can offer better accuracy in computing the statistical properties of stochastic problems, compared to the traditional QMC with equal weights [51, 52]. In this research, these integration formats based on low-discrepancy sample sets assigned non-equal weights are collectively referred to as generalized quasi-MC (GQMC) methods.

Despite there are many successful uncertainty quantification methods mentioned above, it is important to note that these methods are almost entirely non-conditional uncertainty quantification methods. By only considering uncertainties in the structural model and loads, the calculated uncertainties in random structural dynamical responses can sometimes be overly conservative, and hence, may provide limited reference value for the structural safety and reliability analysis. For example, the three-standard deviation interval is often used to measure the uncertainty of response, and is important for the estimation of structural safety and reliability [55]. However, if the range of this three-standard deviation interval is large, it may have limited significance for understanding the structural dynamical behaviors. If in the uncertainty quantification of structural response with random parameters can consider the measurement data, it is clear that the uncertainty of response can be further refined. The main purpose of this research is to investigate how to accurately and efficiently quantify the uncertainty of stochastic linear/nonlinear structural dynamical problems considering measurement conditions with random errors. Essentially, this problem involves estimating the conditional probability density function (PDF), conditional mean, conditional variance, and other conditional statistical properties of the response under specific given conditions, which pertains to conditional uncertainty quantification problems. Most existing uncertainty quantification methods, such as the stochastic perturbation method and the polynomial chaos expansion method, are designed for non-conditional uncertainty quantification problems, and are not yet applicable to solving conditional uncertainty quantification problems.

In this research, a novel key conditional quotient (KCQ) theory and the corresponding numerical method are developed for the conditional uncertainty quantification of linear/nonlinear structural dynamical responses under measurement conditions. The main contributions include: 1) Exact quotient-form expressions for the conditional PDF, conditional mean, and conditional variance of the stochastic structural dynamical response affected by random fields, uncertain parameters, and/or random loads, are proposed. These quotient-form expressions are constructed by extracting key conditions from measurement data and utilizing the principle of probability conservation and the conditional probability theory, and hence, are referred to as key conditional quotients (KCQ). The conditional mean, conditional variance, and conditional PDF of the random response under key measurement conditions are termed KCQ-mean, KCQ-variance, and KCQ-PDF, respectively. When extracting key conditions, the correlation between measurement data and the random response to be quantified is first calculated, followed by the selection of the measurement data with the strongest correlation as the key measurement conditions for the conditional uncertainty quantification of response. 2) Construct an efficient numerical method for calculating KCQs, which employs the GQMC method with non-equal weights to calculate the high-dimensional integrals involved in KCQs, and smooth the Dirac function involved in KCQ-PDF. It also employs an offline-online coupled computational strategy to efficiently calculate the KCQs. In the offline computational stage, the dynamical responses of structures with uncertain parameters sampling in terms of the GQMC are calculated, forming a database that includes response samples, and the corresponding non-equal weights assigned to the samples. In the online computational stage, KCQs of the random response are efficiently calculated, providing the conditional PDF, conditional mean and conditional variance of the random response.

The remainder of the paper is organized as follows: Section 2 provides a detailed description of the conditional uncertainty quantification problem of stochastic linear/nonlinear structural dynamical responses under measurement conditions. Section 3 develops the KCQ theory for the conditional uncertainty quantification under key measurement conditions, as well as the method of extracting key conditions. Section 4 constructs numerical schemes and computational strategy for KCQ-mean, KCQ-variance, and KCQ-PDF using the GQMC and the smoothing technology of the Dirac function. Section 5 provides numerical examples to demonstrate the correctness and effectiveness of the proposed KCQ theory. Section 6 offers a summary of this research.

## 2. Basic statement of the considered problem

Consider a stochastic dynamical finite element equation with structural uncertainty parameters $\boldsymbol{\varepsilon}$ and load uncertainty parameters $\boldsymbol{\theta}$ as follows [3]:

$$\boldsymbol{M}(\boldsymbol{\varepsilon})\ddot{\boldsymbol{u}} + \boldsymbol{C}(\boldsymbol{\varepsilon})\dot{\boldsymbol{u}} + \boldsymbol{F}(\boldsymbol{u},\boldsymbol{\varepsilon}) = \boldsymbol{f}(\boldsymbol{\theta}), \tag{1}$$

where $\boldsymbol{u} \in \mathbb{R}^{N \times 1}$ is the stochastic structural response, $N$ is the number of degrees of freedom of the structure, $\boldsymbol{M}(\boldsymbol{\varepsilon})$ and $\boldsymbol{C}(\boldsymbol{\varepsilon})$ are the mass matrix and damping matrix, respectively, and $\boldsymbol{F}(\boldsymbol{u},\boldsymbol{\varepsilon})$ is the nonlinear restoring force of the structure. If the structure is linear, then $F(\varepsilon, u) = K(\varepsilon)u$, where $K(\varepsilon)$ is the stiffness matrix. The joint probability density function (PDF) of the input random parameters $\varepsilon$ and $\theta$ is known as $\rho_{\varepsilon,\theta}(\varepsilon, \theta)$.

There have been many methods developed to solve Eq. (1), such as the Newmark method, the Wilson method, the precise integration method, and the symplectic Euler midpoint method. Taking the symplectic Euler midpoint method as an example, Eq. (1) can be written as [6]:

$$\begin{cases} s = \dot{u} \\ \dot{s} = M(\varepsilon)^{-1}\left[f(\theta) - C(\varepsilon)s - F(u, \varepsilon)\right] \end{cases}. \tag{2}$$

Let the state vector be $U = \left[u^{\mathrm{T}}, s^{\mathrm{T}}\right]^{\mathrm{T}}$, then Eq. (2) can be rewritten as:

$$\dot{U} = H(U), \quad H(U) = \begin{pmatrix} s \\ M(\varepsilon)^{-1}\left[f(\theta) - C(\varepsilon)s - F(u, \varepsilon)\right] \end{pmatrix}. \tag{3}$$

Discretize time into $t_0, t_1, \cdots, t_k, \cdots, t_{N_{\mathrm{T}}}$, where $N_{\mathrm{T}}$ is the total number of time integration steps. According to the symplectic Euler midpoint method [6], we have

$$U_k = U_{k-1} + H(\bar{U}_k)\Delta t, \quad \bar{U}_k = \frac{U_k + U_{k-1}}{2}, \tag{4}$$

where $\Delta t = t_k - t_{k-1}$ is the time step, and $U_k := U(t_k)$ is the state vector at time $t_k$.

The symplectic Euler midpoint method is an implicit solution scheme that only requires the initial displacement $u_0$ and velocity $s_0$ for the iterative calculation. Combining the initial state vector $U_0 = \left[u_0^{\mathrm{T}}, s_0^{\mathrm{T}}\right]^{\mathrm{T}}$ and Eq. (4), the state vectors at different moments $U_k$ can be obtained. In $U_k$, the first half of the elements are the displacements at time $t_k$, denoted by $u_k := u(t_k)$; and the second half of the elements are the velocities at time $t_k$, denoted by $s_k := s(t_k)$. Once the displacement vector $u_k$ is determined, if one wants to know the response at position $x$ and time $t_k$, denoted by $u_k(x)$, the finite element shape function $N(x)$ can be used to calculate it [56]:

$$u_k(x) = N^T(x) u_k. \tag{5}$$

From Eqs. (3)-(5), it can be seen that $u_k$ is actually a function of the structural uncertainty parameters $\varepsilon$, load uncertainty parameters $\theta$, initial displacement $u_0$, and initial velocity $s_0$, and hence can be denoted by:

$$u_k = g(\varepsilon, \theta, u_0, s_0, t_k). \tag{6}$$

It must be pointed out that, in fact, regardless of which time integration method is used to solve Eq. (1), the displacement $u_k$ can always be written as the function shown in Eq. (6), with the difference being the specific form of the function $g$. In some problems, such as random ocean wave [57] and vibration energy harvester [58], the initial conditions $u_0$ and $s_0$ may also be uncertain, and hence must also be considered in the uncertainty quantification of the response. Here, the random variables $(\varepsilon, \theta, u_0, s_0)$ are uniformly denoted as $\alpha$, thus we have:

$$u_k = g(\alpha, t_k), \quad u_k(x) = N^T(x) g(\alpha, t_k) := g(\alpha, t_k, x). \tag{7}$$

In many practical engineering applications, researchers often set some measurement points at some crucial locations of the structure to measure the response values. However, it is inevitable to have measurement errors in the actual measurement process, especially the structure is in vibration [58]. Considering random measurement errors, we have

$$y(t) = h(u(t)) + v(t) = h(g(\varepsilon, \theta, u_0, s_0, t)) + v(t), \tag{8}$$

where $y(t) \in \mathbb{R}^{N_m \times 1}$ is the real measurement value at time $t$, $v(t) \in \mathbb{R}^{N_m \times 1}$ is the random measurement error of $y(t)$, $h(\cdot)$ represents the measurement equation, describing how the response is measured, and $N_m$ is the number of measurement points. Each measurement point can return the measurement value of the structural response at different times $t = [t_1, t_2, \cdots, t_k]$, and for convenience of analysis, let $\mathbf{y}_k = \mathbf{y}(t_k)$, and $\mathbf{v}_k = \mathbf{v}(t_k)$. Considering both the dynamical equation (1) and the measurement equation (8), we have:

$$\begin{cases} \mathbf{u}_k = \mathbf{g}(\boldsymbol{\alpha}, t_k) \\ \mathbf{y}_k = \mathbf{h}(\mathbf{u}_k) + \mathbf{v}_k \\ \vdots \\ \mathbf{y}_1 = \mathbf{h}(\mathbf{u}_1) + \mathbf{v}_1 \end{cases}, \qquad (9)$$

where $\mathbf{y}_k = (y_{k,1}, y_{k,2}, \cdots, y_{k,N_m})^\mathrm{T}$, and $y_{k,i}$ is the data measured by the $i$-th measurement point at time $t_k$. Equation (9) takes into account both the stochastic dynamical equation and the stochastic measurement equations. The main issue of this report is to study how to quantify the uncertainty of the structural random response under a series of measurement conditions $\mathbf{y}_1, \mathbf{y}_2, \cdots, \mathbf{y}_k$ based on Eq. (9). Specifically, it includes the quantifications of conditional PDF, conditional mean, and conditional variance.

In this study, it is assumed that the joint PDF $\rho_{\varepsilon,\theta}(\varepsilon, \theta)$, and the PDF $\rho_v(\mathbf{v})$ of the random measurement error are known, and $(\varepsilon, \theta)$ and $\mathbf{v}$ are independent of each other. To quantify the uncertainty of $\mathbf{u}_k$ under the measurement conditions $\mathbf{y}_{1:k} = [\mathbf{y}_1^\mathrm{T}, \mathbf{y}_2^\mathrm{T}, \cdots, \mathbf{y}_k^\mathrm{T}]^\mathrm{T}$, it is needed to use the conditional PDF $\rho(\mathbf{u}_k | \mathbf{y}_{1:k})$ which can be expressed as [59]:

$$\rho_{\mathbf{u}_k | \mathbf{y}_{1:k}}(\mathbf{u}_k | \mathbf{y}_{1:k}) = \frac{\rho_{\mathbf{u}_k, \mathbf{y}_{1:k}}(\mathbf{u}_k, \mathbf{y}_{1:k})}{\rho_{\mathbf{y}_{1:k}}(\mathbf{y}_{1:k})}, \qquad (10)$$

where $\rho_{\mathbf{u}_k, \mathbf{y}_{1:k}}(\mathbf{u}_k, \mathbf{y}_{1:k})$ is the joint PDF of $\mathbf{u}_k$ and $\mathbf{y}_{1:k}$, and $\rho_{\mathbf{y}_{1:k}}(\mathbf{y}_{1:k})$ is the PDF of $\mathbf{y}_{1:k}$. According to the conditional PDF $\rho(\mathbf{u}_k | \mathbf{y}_{1:k})$, the conditional mean and conditional covariance of $\mathbf{u}_k$ considering the measurement data $\mathbf{y}_{1:k}$ can be expressed as [59]:

$$E(\boldsymbol{u}_k \mid \boldsymbol{y}_{1:k}) = \int_{-\infty}^{+\infty} \rho(\boldsymbol{u}_k \mid \boldsymbol{y}_{1:k}) \boldsymbol{u}_k \mathrm{d}\boldsymbol{u}_k = \hat{\boldsymbol{u}}_k, \tag{11}$$

and

$$E\left((\boldsymbol{u}_k - \hat{\boldsymbol{u}}_k)(\boldsymbol{u}_k - \hat{\boldsymbol{u}}_k)^{\mathrm{T}} \mid \boldsymbol{y}_{1:k}\right) = \int_{-\infty}^{+\infty} \rho(\boldsymbol{u}_k \mid \boldsymbol{y}_{1:k})(\boldsymbol{u}_k - \hat{\boldsymbol{u}}_k)(\boldsymbol{u}_k - \hat{\boldsymbol{u}}_k)^{\mathrm{T}} \mathrm{d}\boldsymbol{u}_k = \boldsymbol{P}_k. \tag{12}$$

Equations (10)-(12) are the conditional PDF, conditional mean, and conditional covariance of the response considering the measurement data. They are the statistical quantities that need to be calculated in this research. It can be observed from Eqs. (10)-(12), this research studies the conditional uncertainty quantification, while the traditional uncertainty quantification is non-conditional. Hence, it is necessary to develop new analytical and numerical methods to quantify conditional uncertainty. In the next section, based on the principle of probability conservation (see Appendix), the exact expressions for the conditional PDF, conditional mean, and conditional covariance of the random response will be derived.

## 3. Key conditional quotient theory

### 3.1. Uncertainty Quantification of stochastic dynamical response considering measurement conditions

For ease of analysis, the measurement errors at a series of discrete moments $\boldsymbol{t} = [t_1, t_2, \cdots, t_k]$ are denoted as $\boldsymbol{v}_{1:k} = \left[\boldsymbol{v}_1^{\mathrm{T}}, \boldsymbol{v}_2^{\mathrm{T}}, \cdots, \boldsymbol{v}_k^{\mathrm{T}}\right]^{\mathrm{T}}$, and the joint PDF of the random variables $\boldsymbol{u}_k$, $\boldsymbol{y}_{1:k}$, $\boldsymbol{v}_{1:k}$, and $\boldsymbol{\alpha}$ is denoted as $\rho_{\boldsymbol{u}_k, \boldsymbol{y}_{1:k}, \boldsymbol{v}_{1:k}, \boldsymbol{\alpha}}(\boldsymbol{u}_k, \boldsymbol{y}_{1:k}, \boldsymbol{v}_{1:k}, \boldsymbol{\alpha})$. According to the principle of probability conservation (see Appendix), it is known that

$$\rho_{\boldsymbol{u}_k, \boldsymbol{y}_{1:k}, \boldsymbol{v}_{1:k}, \boldsymbol{\alpha}}(\boldsymbol{u}_k, \boldsymbol{y}_{1:k}, \boldsymbol{v}_{1:k}, \boldsymbol{\alpha}) = \rho_{\boldsymbol{\alpha}}(\boldsymbol{\alpha}) \rho_{\boldsymbol{v}_1, \cdots, \boldsymbol{v}_k}(\boldsymbol{v}_1, \cdots, \boldsymbol{v}_k) \delta(\boldsymbol{u}_k - \boldsymbol{g}(\boldsymbol{\alpha}, t_k)) \\ \times \prod_{i=1}^{k} \delta(\boldsymbol{y}_i - \boldsymbol{h}(\boldsymbol{g}(\boldsymbol{\alpha}, t_i)) - \boldsymbol{v}_i) \tag{13}$$

By integrating the above equation, the marginal PDFs $\rho_{\boldsymbol{u}_k, \boldsymbol{y}_{1:k}}(\boldsymbol{u}_k, \boldsymbol{y}_{1:k})$ and $\rho_{\boldsymbol{y}_{1:k}}(\boldsymbol{y}_{1:k})$ can be obtained as follows:

$$\begin{aligned}
\rho_{u_k,y_{1:k}}(u_k, y_{1:k}) &= \int_{-\infty}^{+\infty} \rho_{u_k,y_{1:k},v,\alpha}(u_k, y_{1:k}, v_{1:k}, \alpha) dv_{1:k} d\alpha \\
&= \int_{-\infty}^{+\infty} \begin{bmatrix} \rho_\alpha(\alpha) \rho_{v_1,\cdots,v_k}(v_1,\cdots,v_k) \delta(u_k - g(\alpha,t_k)) \\ \times \prod_{i=1}^{k} \delta(y_i - h(g(\alpha,t_i)) - v_i) \end{bmatrix} dv_1,\cdots,dv_k d\alpha \\
&= \int_{-\infty}^{+\infty} \rho_\alpha(\alpha) \rho_{v_1,\cdots,v_k}(y_1 - h(g(\alpha,t_1)),\cdots, y_k - h(g(\alpha,t_k))) \delta(u_k - g(\alpha,t_k)) d\alpha
\end{aligned} \quad (14)$$

and

$$\begin{aligned}
\rho_{y_{1:k}}(y_{1:k}) &= \int_{-\infty}^{+\infty} \rho_{u_k,y_{1:k},v,\alpha}(u_k, y_{1:k}, v_{1:k}, \alpha) dv_{1:k} d\alpha du_k \\
&= \int_{-\infty}^{+\infty} \rho_\alpha(\alpha) \rho_{v_1,\cdots,v_k}(y_1 - h(g(\alpha,t_1)),\cdots, y_k - h(g(\alpha,t_k))) \delta(u_k - g(\alpha,t_k)) du_k d\alpha ,(15) \\
&= \int_{-\infty}^{+\infty} \rho_\alpha(\alpha) \rho_{v_1,\cdots,v_k}(y_1 - h(g(\alpha,t_1)),\cdots, y_k - h(g(\alpha,t_k))) d\alpha
\end{aligned}$$

where the integral $\int_{-\infty}^{+\infty} (\bullet) dv_1,\cdots,dv_k d\alpha$ represents the multiple integral $\int_{-\infty}^{+\infty} \cdots \int_{-\infty}^{+\infty} (\bullet) dv_1,\cdots,dv_k d\alpha$. Substituting Eqs. (14) and (15) into Eq. (10), the exact expression for the conditional PDF $\rho(u_k | y_{1:k})$ can be derived:

$$\rho(u_k | y_{1:k}) = \frac{\int_{-\infty}^{+\infty} \rho_\alpha(\alpha) \rho_{v_1,\cdots,v_k}(y_1 - h(g(\alpha,t_1)),\cdots, y_k - h(g(\alpha,t_k))) \delta(u_k - g(\alpha,t_k)) d\alpha}{\int_{-\infty}^{+\infty} \rho_\alpha(\alpha) \rho_{v_1,\cdots,v_k}(y_1 - h(g(\alpha,t_1)),\cdots, y_k - h(g(\alpha,t_k))) d\alpha} . \quad (16)$$

Further substituting Eq. (16) into Eqs.(11) and (12) yields:

$$\begin{aligned}
\hat{u}_k &= \int_{-\infty}^{+\infty} \rho(u_k | y_{1:k}) u_k du_k \\
&= \frac{\int_{-\infty}^{+\infty} \rho_\alpha(\alpha) \rho_{v_1,\cdots,v_k}(y_1 - h(g(\alpha,t_1)),\cdots, y_k - h(g(\alpha,t_k))) \delta(u_k - g(\alpha,t_k)) u_k du_k d\alpha}{\int_{-\infty}^{+\infty} \rho_\alpha(\alpha) \rho_{v_1,\cdots,v_k}(y_1 - h(g(\alpha,t_1)),\cdots, y_k - h(g(\alpha,t_k))) d\alpha} \\
&= \frac{\int_{-\infty}^{+\infty} \rho_\alpha(\alpha) \rho_{v_1,\cdots,v_k}(y_1 - h(g(\alpha,t_1)),\cdots, y_k - h(g(\alpha,t_k))) g(\alpha,t_k) d\alpha}{\int_{-\infty}^{+\infty} \rho_\alpha(\alpha) \rho_{v_1,\cdots,v_k}(y_1 - h(g(\alpha,t_1)),\cdots, y_k - h(g(\alpha,t_k))) d\alpha}
\end{aligned} \quad (17)$$

and

$$P_k = \int_{-\infty}^{+\infty} \rho(\boldsymbol{u}_k \mid \boldsymbol{y}_{1:k})(\boldsymbol{u}_k - \hat{\boldsymbol{u}}_k)(\boldsymbol{u}_k - \hat{\boldsymbol{u}}_k)^\mathrm{T} \mathrm{d}\boldsymbol{u}_k$$

$$= \frac{\int_{-\infty}^{+\infty} \left[ \begin{array}{l} \rho_\alpha(\boldsymbol{\alpha}) \rho_{v_1,\cdots,v_k}\left(\boldsymbol{y}_1 - \boldsymbol{h}(\boldsymbol{g}(\boldsymbol{\alpha},t_1)),\cdots,\boldsymbol{y}_k - \boldsymbol{h}(\boldsymbol{g}(\boldsymbol{\alpha},t_k))\right) \\ \times \delta(\boldsymbol{u}_k - \boldsymbol{g}(\boldsymbol{\alpha},t_k))(\boldsymbol{u}_k - \hat{\boldsymbol{u}}_k)(\boldsymbol{u}_k - \hat{\boldsymbol{u}}_k)^\mathrm{T} \end{array}\right] \mathrm{d}\boldsymbol{u}_k \mathrm{d}\boldsymbol{\alpha}}{\int_{-\infty}^{+\infty} \rho_\alpha(\boldsymbol{\alpha}) \rho_{v_1,\cdots,v_k}\left(\boldsymbol{y}_1 - \boldsymbol{h}(\boldsymbol{g}(\boldsymbol{\alpha},t_1)),\cdots,\boldsymbol{y}_k - \boldsymbol{h}(\boldsymbol{g}(\boldsymbol{\alpha},t_k))\right) \mathrm{d}\boldsymbol{\alpha}}$$

$$= \frac{\int_{-\infty}^{+\infty} \left[ \begin{array}{l} \rho_\alpha(\boldsymbol{\alpha}) \rho_{v_1,\cdots,v_k}\left(\boldsymbol{y}_1 - \boldsymbol{h}(\boldsymbol{g}(\boldsymbol{\alpha},t_1)),\cdots,\boldsymbol{y}_k - \boldsymbol{h}(\boldsymbol{g}(\boldsymbol{\alpha},t_k))\right) \\ \times (\boldsymbol{g}(\boldsymbol{\alpha},t_k) - \hat{\boldsymbol{u}}_k)(\boldsymbol{g}(\boldsymbol{\alpha},t_k) - \hat{\boldsymbol{u}}_k)^\mathrm{T} \end{array}\right] \mathrm{d}\boldsymbol{\alpha}}{\int_{-\infty}^{+\infty} \rho_\alpha(\boldsymbol{\alpha}) \rho_{v_1,\cdots,v_k}\left(\boldsymbol{y}_1 - \boldsymbol{h}(\boldsymbol{g}(\boldsymbol{\alpha},t_1)),\cdots,\boldsymbol{y}_k - \boldsymbol{h}(\boldsymbol{g}(\boldsymbol{\alpha},t_k))\right) \mathrm{d}\boldsymbol{\alpha}}. \quad (18)$$

Equations (16), (17) and (18) represent the exact expressions for the conditional PDF, conditional mean, and conditional covariance of the random response $\boldsymbol{u}_k$, given the measurement data $\boldsymbol{y}_{1:k}$. It is observed that the expressions for the conditional PDF, conditional mean and conditional covariance derived above are all in the quotient-form, henceforth referred to as the conditional quotients (CQs) in the subsequent text. The denominators of the three CQs are the same, and in the denominator, there exists a joint PDF of random measurement errors at different moments, i.e., $\rho_{v_1,\cdots,v_k}\left(\boldsymbol{y}_1 - \boldsymbol{h}(\boldsymbol{g}(\boldsymbol{\alpha},t_1)),\cdots,\boldsymbol{y}_k - \boldsymbol{h}(\boldsymbol{g}(\boldsymbol{\alpha},t_k))\right)$. As $k$ increases, the value of this joint PDF may gradually approach zero, leading to the failure of numerical calculations for the CQs. To further illustrate this situation, we refer to the event "the measurement error $\boldsymbol{v}_i$ at time $t_i$ is exactly $\boldsymbol{y}_i - \boldsymbol{h}(\boldsymbol{g}(\boldsymbol{\alpha},t_i))$" as "the measurement event $i$" ( denoted by ME $i$ ). Therefore, the value of $\rho_{v_1,\cdots,v_k}\left(\boldsymbol{y}_1 - \boldsymbol{h}(\boldsymbol{g}(\boldsymbol{\alpha},t_1)),\cdots,\boldsymbol{y}_k - \boldsymbol{h}(\boldsymbol{g}(\boldsymbol{\alpha},t_k))\right)$ describes the likelihood of MEs 1-$k$ occurring simultaneously. The greater the likelihood of these $k$ MEs occurring simultaneously, the larger the value of $\rho_{v_1,\cdots,v_k}\left(\boldsymbol{y}_1 - \boldsymbol{h}(\boldsymbol{g}(\boldsymbol{\alpha},t_1)),\cdots,\boldsymbol{y}_k - \boldsymbol{h}(\boldsymbol{g}(\boldsymbol{\alpha},t_k))\right)$, and vice versa. Obviously, from a probabilistic perspective, the likelihood of multiple events occurring simultaneously is generally less than that of a single event. In the CQs (16)-(18), as $k$ increases, the likelihood of these $k$ MEs occurring simultaneously will become increasingly smaller, that is the value of $\rho_{v_1,\cdots,v_k}\left(y_1-h(g(\alpha,t_1)),\cdots,y_k-h(g(\alpha,t_k))\right)$ will also become smaller and smaller. How to address the issue of $\rho_{v_1,\cdots,v_k}\left(y_1-h(g(\alpha,t_1)),\cdots,y_k-h(g(\alpha,t_k))\right)$ becoming smaller and smaller, causing the denominator in the CQs to approach zero, and potentially leading to numerical failure, will be further discussed in the next subsection.

**3.2. Uncertainty Quantification of stochastic dynamical response considering only the key measurement conditions**

The analysis in subsection 3.1 indicates that the joint PDF of random measurement errors, i.e., $\rho_{v_1,\cdots,v_k}\left(y_1-h(g(\alpha,t_1)),\cdots,y_k-h(g(\alpha,t_k))\right)$ which appears in the denominators of CQs, tends to approach zero as the number of time steps $k$ increases, leading to the failure of the numerical calculation. Actually, the derivation of CQs (16)-(18) uses implicitly an assumption that all measurement data have equal reference value for the uncertainty quantification of the stochastic dynamical response. In real engineering applications, multiple measurement points are installed at different locations to measurement the dynamical response values at different moments, resulting in multiple sets of measurement values, denoted as $y_{1:k}=\left[y_1^T,y_2^T,\cdots,y_k^T\right]^T$. However, for the uncertainty quantification analysis of the stochastic response $u_k(x)$ at a specific location $x$, it is clear that not all measurement data have the same reference value. For example, for the uncertainty quantification of the response $u_k(x)$ at time $t_k$, the reference value of the measurement data $y_1$ at initial time is obviously not as key as the reference value of the measurement data $y_k$ at time $t_k$; similarly, the reference value of the measurement data form measurement points far from $x$ is likely not as key as that from measurement points closer to $x$. Therefore, we can extract the measurement conditions with the greatest reference value, i.e., key conditions, to quantify the uncertainty of $u_k(x)$, which is the starting point of this research.

We first investigate how to extract key measurement conditions from all $y_{1:k}$ which contains measurement data from different measurement points at different moments. And it is needed to select the $N_k$ measurement data that have the most key reference value for the uncertainty quantification analysis of the random response $u_k(x)$ at time $t_k$ and location $x$. To this end, we first evaluate the correlation coefficient between the response $u_k(x)$ and the measurement data $y_{i,j}$ obtained by the $j$-th measurement point at time $t_i$, and the formula for the calculation of the correlation coefficient is as follows:

$$r_{i,j,k}(x) = \frac{\text{cov}(y_{i,j}, u_k(x))}{\sqrt{\text{cov}(y_{i,j}, y_{i,j})}\sqrt{\text{cov}(u_k(x), u_k(x))}}. \tag{19}$$

From Eq. (19), it is easy to see that $|r_{i,j,k}(x)|$ can be used to assess the degree of correlation between the measurement data $y_{i,j}$ and the response $u_k(x)$. It is believed that the greater the degree of correlation $|r_{i,j,k}(x)|$, the more important the reference value of $y_{i,j}$ for the uncertainty quantification analysis of $u_k(x)$. Therefore, we select the $N_k$ strong correlation measurement data with the largest $|r_{i,j,k}(x)|$ as key measurement conditions to quantify the uncertainty of $u_k(x)$, and ignore other measurement data with smaller correlation coefficients. This is the core idea for extracting key measurement conditions.

Next, we discuss how to quantify the uncertainty of $u_k(x)$ based on the key measurement conditions. For the sake of discussion, let the vector of key measurement data selected according to $|r_{i,j,k}(x)|$ be $z_k(x)$. The relationship between the key measurement data $z_k(x)$ and the total measurement data $y_{1:k}$ can be expressed as:

$$z_k(x) = p_k(x) y_{1:k}, \tag{20}$$

where $p_k(x)$ is the key condition matrix with $N_k$ rows and $N_m \times k$ columns, and all matrix elements are 0 or 1. Using the key condition matrix, the stochastic observation equation (9) can be rewritten as:

$$z_k = \hat{h}_k(\alpha, x) + \beta_k, \tag{21}$$

where

$$\hat{h}_k(\alpha, x) = p_k(x) \begin{bmatrix} h(u_1) \\ \vdots \\ h(u_k) \end{bmatrix} = p_k(x) \begin{bmatrix} h(g(\alpha, t_1)) \\ \vdots \\ h(g(\alpha, t_k)) \end{bmatrix}, \quad \beta_k = p_k(x) \begin{bmatrix} v_1 \\ \vdots \\ v_k \end{bmatrix} = p_k(x) v_{1:k}, \tag{22}$$

and $\beta_k$ represents the random error vectors of key measurement data.

Combining Eqs. (7) and (21) yields the stochastic analysis model considering key measurement conditions $z_k(x)$ to quantify the uncertainty of the stochastic dynamical response $u_k(x)$ at location $x$:

$$\begin{cases} u_k(x) = g(\alpha, t_k, x) \\ z_k = \hat{h}_k(\alpha, x) + \beta_k \end{cases}. \tag{23}$$

Based on Eq. (23), we can derive the condition mean, conditional variance, and conditional PDF of the stochastic response $u_k(x)$ considering key measurement conditions $z_k(x)$. Using the same derivation process as in subsection 3.1, we can obtain the following joint PDF:

$$\rho_{u_k, z_k, \beta_k, \alpha}(u_k, z_k, \beta_k, \alpha) = \rho_\alpha(\alpha) \rho_{\beta_k}(\beta_k) \delta(u_k - g(\alpha, t_k, x)) \delta(z_k - \hat{h}_k(\alpha, x) - \beta_k), \tag{24}$$

where $\rho_{\beta_k}(\beta_k)$ represent the joint PDF of $\beta_k$. According to Eq. (24), we can obtain the marginal probability densities $\rho_{u_k, z_k}(u_k, z_k)$ and $\rho_{z_k}(z_k)$:

$$\rho_{u_k,z_k}(u_k,z_k) = \int_{-\infty}^{+\infty} \rho_{u_k,z_k,\beta_k,\alpha}(u_k,z_k,\beta_k,\alpha)\,\mathrm{d}\beta_k\mathrm{d}\alpha$$
$$= \int_{-\infty}^{+\infty}\left[\rho_\alpha(\alpha)\rho_{\beta_k}(\beta_k)\delta(u_k - g(\alpha,t_k,x))\delta(z_k - \hat{h}_k(\alpha,x) - \beta_k)\right]\mathrm{d}\beta_k\mathrm{d}\alpha, \quad (25)$$
$$= \int_{-\infty}^{+\infty}\left[\rho_\alpha(\alpha)\rho_{\beta_k}(z_k - \hat{h}_k(\alpha,x))\delta(u_k - g(\alpha,t_k,x))\right]\mathrm{d}\alpha$$

and

$$\rho_{z_k}(z_k) = \int_{-\infty}^{+\infty}\rho_{u_k,z_k}(u_k,z_k)\mathrm{d}u_k$$
$$= \int_{-\infty}^{+\infty}\int_{-\infty}^{+\infty}\left[\rho_\alpha(\alpha)\rho_{\beta_k}(z_k - \hat{h}_k(\alpha,x))\delta(u_k - g(\alpha,t_k,x))\right]\mathrm{d}\alpha\mathrm{d}u_k. \quad (26)$$
$$= \int_{-\infty}^{+\infty}\rho_\alpha(\alpha)\rho_{\beta_k}(z_k - \hat{h}_k(\alpha,x))\mathrm{d}\alpha$$

Using Eqs. (25) and (26), the conditional PDF $\rho_{u_k|z_k}(u_k|z_k)$ considering key conditions can be obtained as follows:

$$\rho_{u_k|z_k}(u_k|z_k) = \frac{\int_{-\infty}^{+\infty}\left[\rho_\alpha(\alpha)\rho_{\beta_k}(z_k - \hat{h}_k(\alpha,x))\delta(u_k - g(\alpha,t_k,x))\right]\mathrm{d}\alpha}{\int_{-\infty}^{+\infty}\rho_\alpha(\alpha)\rho_{\beta_k}(z_k - \hat{h}_k(\alpha,x))\mathrm{d}\alpha}. \quad (27)$$

According to the conditional PDF shown in Eq. (27), the conditional mean and conditional covariance of $u_k(x)$ considering key conditions $z_k$ can be written as:

$$\hat{u}_k(x) = \int_{-\infty}^{+\infty}\rho_{u_k|z_k}(u_k|z_k)u_k\mathrm{d}u_k = \frac{\int_{-\infty}^{+\infty}\left[\rho_\alpha(\alpha)\rho_{\beta_k}(z_k - \hat{h}_k(\alpha,x))g(\alpha,t_k,x)\right]\mathrm{d}\alpha}{\int_{-\infty}^{+\infty}\rho_\alpha(\alpha)\rho_{\beta_k}(z_k - \hat{h}_k(\alpha,x))\mathrm{d}\alpha}, \quad (28)$$

and

$$P_k(x) = \int_{-\infty}^{+\infty}\rho_{u_k|z_k}(u_k|z_k)(u_k - \hat{u}_k)(u_k - \hat{u}_k)^\mathrm{T}\mathrm{d}u_k$$
$$= \frac{\int_{-\infty}^{+\infty}\left[\rho_\alpha(\alpha)\rho_{\beta_k}(z_k - \hat{h}_k(\alpha,x))(g(\alpha,t_k,x) - \hat{u}_k)(g(\alpha,t_k,x) - \hat{u}_k)^\mathrm{T}\right]\mathrm{d}\alpha}{\int_{-\infty}^{+\infty}\rho_\alpha(\alpha)\rho_{\beta_k}(z_k - \hat{h}_k(\alpha,x))\mathrm{d}\alpha}. \quad (29)$$

Equations (27)-(29) are the conditional PDF, conditional mean, and conditional variance of the stochastic response $u_k(x)$ considering key measurement conditions $z_k$. It can be seen that these three statistical quantities considering key conditions are all in the quotient-form, henceforth referred to as the key conditional quotient (KCQ) in this paper. When specially referring to a statistical quantity # that considers key conditions, it will be referred to as KCQ-#. For example, the conditional PDF of the response considering key conditions will be referred to as KCQ-PDF. The above analysis constructs the KCQs for uncertainty quantification of the stochastic displacement response $u_k(\boldsymbol{x})$. In fact, for other stochastic responses such as velocity and stress, their KCQs can be derived by using the same approach, and the derivation process is not repeated here.

## 4. Numerical method proposed for KCQ

**4.1 Numerical schemes for KCQ-mean and KCQ-variance**

It can be observed that in the KCQs shown in Eqs. (27)-(29), both the numerator and the denominator involve high-dimensional integrals which are difficult to solve analytically. GQMC-type algorithms proposed in recent years have been proven to be an accurate and effective algorithm for solving high-dimensional numerical integration problems. The GQMC first proposed by Chen and Li et al. [51, 52] utilizes low-discrepancy samples as integration points and assigns corresponding non-equal weights to each integration point through Voronoi diagrams. Subsequently, Wu and Zhao et al. [54] proposed the generalized L2 discrepancy based on general points, and provided a generation method for non-equal weights based on the principle of minimum discrepancy. At present, due to the accurate and efficient advantages, the GQMC method has been widely used in the research of uncertainty quantification [36, 37]. In this paper the GQMC are introduced to deal with high-dimensional integrals involved in KCQs.

Based on the GQMC method, any high-dimensional integral can be approximated as:

$$I = \int_{-\infty}^{+\infty} \rho_{\boldsymbol{\alpha}}(\boldsymbol{\alpha}) f(\boldsymbol{\alpha}) \mathrm{d}\boldsymbol{\alpha} \approx \sum_{i=1}^{n} w_i f(\boldsymbol{\alpha}_i), \tag{30}$$

where $f(\boldsymbol{\alpha})$ is the considered random function, $\boldsymbol{\alpha}_i$ represents the $i$-th sample, $w_i$ represents the weight of the $i$-th sample point, and $n$ is the number of samples. For the generation methods of sample points and non-equal weights, interested readers can refer to the work of Chen and Li et al. [51, 52], as well as the work of Wu and Zhao et al. [54].

Using the GQMC method to calculate the KCQ-mean and KCQ-variance shown in Eqs. (28) and (29). First, generate $n$ sample points $\boldsymbol{\alpha}_i$ and the corresponding non-equal weights $w_i$, $i = 1, 2, \cdots, n$, according to the GQMC and the joint PDF $\rho_\alpha(\boldsymbol{\alpha})$. Then substitute these $n$ sample points $\boldsymbol{\alpha}_i$ into the stochastic dynamical equation (1). Solving the stochastic dynamical equation yields $n$ sets of $g(\boldsymbol{\alpha}_i, t_k, \boldsymbol{x})$ and $\hat{\boldsymbol{h}}_k(\boldsymbol{\alpha}_i, \boldsymbol{x})$. Once $g(\boldsymbol{\alpha}_i, t_k, \boldsymbol{x})$ and $\hat{\boldsymbol{h}}_k(\boldsymbol{\alpha}_i, \boldsymbol{x})$ are obtained, in terms of Eq. (30), we have

$$\int_{-\infty}^{+\infty} \rho_\alpha(\boldsymbol{\alpha}) \rho_{\beta_k}\left(z_k - \hat{\boldsymbol{h}}_k(\boldsymbol{\alpha}, \boldsymbol{x})\right) d\boldsymbol{\alpha} \approx \sum_{i=1}^{n} w_i \rho_{\beta_k}\left(z_k - \hat{\boldsymbol{h}}_k(\boldsymbol{\alpha}_i, \boldsymbol{x})\right), \tag{31}$$

$$\int_{-\infty}^{+\infty} \left[\rho_\alpha(\boldsymbol{\alpha}) \rho_{\beta_k}\left(z_k - \hat{\boldsymbol{h}}_k(\boldsymbol{\alpha}, \boldsymbol{x})\right) g(\boldsymbol{\alpha}_i, t_k, \boldsymbol{x})\right] d\boldsymbol{\alpha} \approx \sum_{i=1}^{n} w_i \left[\rho_{\beta_k}\left(z_k - \hat{\boldsymbol{h}}_k(\boldsymbol{\alpha}_i, \boldsymbol{x})\right) g(\boldsymbol{\alpha}_i, t_k, \boldsymbol{x})\right], \tag{32}$$

and

$$\begin{aligned}
&\int_{-\infty}^{+\infty} \left[\begin{array}{l} \rho_\alpha(\boldsymbol{\alpha}) \rho_{\beta_k}\left(z_k - \hat{\boldsymbol{h}}_k(\boldsymbol{\alpha}, \boldsymbol{x})\right) \\ \times \left(g(\boldsymbol{\alpha}_i, t_k, \boldsymbol{x}) - \hat{u}_k\right)\left(g(\boldsymbol{\alpha}_i, t_k, \boldsymbol{x}) - \hat{u}_k\right)^{\mathrm{T}} \end{array}\right] d\boldsymbol{\alpha} \\
&\approx \sum_{i=1}^{n} w_i \left[\begin{array}{l} \rho_{\beta_k}\left(z_k - \hat{\boldsymbol{h}}_k(\boldsymbol{\alpha}_i, \boldsymbol{x})\right)\left(g(\boldsymbol{\alpha}_i, t_k, \boldsymbol{x}) - \hat{u}_k\right) \\ \times \left(g(\boldsymbol{\alpha}_i, t_k, \boldsymbol{x}) - \hat{u}_k\right)^{\mathrm{T}} \end{array}\right].
\end{aligned} \tag{33}$$

Substituting Eqs. (31)-(33) into Eqs. (28) and (29), KCQ-mean and KCQ-variance can be approximated as

$$\hat{u}_k(x) = \frac{\sum_{i=1}^{n} w_i \left[ \rho_{\beta_k}\left(z_k - \hat{h}_k(\alpha_i, x)\right) g(\alpha_i, t_k, x) \right]}{\sum_{i=1}^{n} w_i \rho_{\beta_k}\left(z_k - \hat{h}_k(\alpha_i, x)\right)}, \tag{34}$$

and

$$P_k(x) = \frac{\sum_{i=1}^{n} w_i \left[ \rho_{\beta_k}\left(z_k - \hat{h}_k(\alpha_i, x)\right) \left(g(\alpha_i, t_k, x) - \hat{u}_k\right)\left(g(\alpha_i, t_k, x) - \hat{u}_k\right)^{\mathrm{T}} \right]}{\sum_{i=1}^{n} w_i \rho_{\beta_k}\left(z_k - \hat{h}_k(\alpha_i, x)\right)}, \tag{35}$$

respectively. The above two formulas provide the numerical schemes for KCQ-mean and KCQ-variance.

Next, we discuss the calculation of $\rho_{\beta_k}(\boldsymbol{\beta})$. In many practical engineering problems, measurement information is often obtained by averaging multiple measurements. According to the central limit theorem [59], the random error of the averaged measurement is usually normally distributed. Hence, the PDF $\rho_{\beta_k}(\boldsymbol{\beta})$ here is assumed to be Gaussian. For the dynamical problem, the measurement error is regarded as a Gaussian random process [60]. Record all the measurement errors at all times as $\boldsymbol{v}_{1:k} = \left[\boldsymbol{v}_1^{\mathrm{T}}, \boldsymbol{v}_2^{\mathrm{T}}, \cdots, \boldsymbol{v}_k^{\mathrm{T}}\right]^{\mathrm{T}}$. Since $\boldsymbol{v}_{1:k}$ is a Gaussian random process, according to Eq. (22), $\boldsymbol{\beta}_k = \boldsymbol{p}_k(x)\boldsymbol{v}_{1:k}$ also follows a Gaussian distribution [59]. Let the mean of the measurement error $\boldsymbol{v}_k$ be $\boldsymbol{\mu}_{v,k}$, the mean and covariance matrix of all the measurement errors at different times be $\boldsymbol{\mu}_{v,1:k} = \left[\boldsymbol{\mu}_{v,1}^{\mathrm{T}}, \boldsymbol{\mu}_{v,2}^{\mathrm{T}}, \cdots, \boldsymbol{\mu}_{v,k}^{\mathrm{T}}\right]^{\mathrm{T}}$ and $\boldsymbol{R}_{v,1:k}$, respectively, then the mean and covariance matrix of $\boldsymbol{\beta}_k = \boldsymbol{p}_k(x)\boldsymbol{v}_{1:k}$ can be expressed as

$$\boldsymbol{\mu}_{\beta_k} = E(\boldsymbol{\beta}_k) = E(\boldsymbol{p}_k(x)\boldsymbol{v}_{1:k}) = \boldsymbol{p}_k(x)\boldsymbol{\mu}_{v,1:k}, \tag{36}$$

and

$$\begin{aligned}
\boldsymbol{R}_{\beta_k} &= E\left(\left(\boldsymbol{\beta}_k - \boldsymbol{\mu}_{\beta_k}\right)\left(\boldsymbol{\beta}_k - \boldsymbol{\mu}_{\beta_k}\right)^{\mathrm{T}}\right) \\
&= E\left(\boldsymbol{p}_k(\boldsymbol{x})\left(\boldsymbol{v}_{1:k} - \boldsymbol{\mu}_{v,1:k}\right)\left(\boldsymbol{v}_{1:k} - \boldsymbol{\mu}_{v,1:k}\right)^{\mathrm{T}} \boldsymbol{p}_k^{\mathrm{T}}(\boldsymbol{x})\right) \\
&= \boldsymbol{p}_k(\boldsymbol{x}) E\left(\left(\boldsymbol{v}_{1:k} - \boldsymbol{\mu}_{v,1:k}\right)\left(\boldsymbol{v}_{1:k} - \boldsymbol{\mu}_{v,1:k}\right)^{\mathrm{T}}\right) \boldsymbol{p}_k^{\mathrm{T}}(\boldsymbol{x}) \\
&= \boldsymbol{p}_k(\boldsymbol{x}) \boldsymbol{R}_{v,1:k} \boldsymbol{p}_k^{\mathrm{T}}(\boldsymbol{x})
\end{aligned} \tag{37}$$

and $\rho_{\beta_k}(\boldsymbol{\beta})$ can be expressed as

$$\rho_{\beta_k}(\boldsymbol{\beta}_k) = \frac{1}{\sqrt{(2\pi)^{N_k} |\boldsymbol{R}_{\beta_k}|}} \exp\left(-\frac{1}{2}\left(\boldsymbol{\beta}_k - \boldsymbol{\mu}_{\beta_k}\right)^{\mathrm{T}} \boldsymbol{R}_{\beta_k}^{-1}\left(\boldsymbol{\beta}_k - \boldsymbol{\mu}_{\beta_k}\right)\right). \tag{38}$$

Using Eqs. (34), (35), and (38), KCQ-mean and KCQ-variance can be calculated.

**4.2 Numerical scheme for the KCQ-PDF**

It can be seen from the KCQ-PDF shown in Eq. (27) that the numerator is also a high-dimensional integral containing a non-smooth Dirac function $\delta(u_k - g(\boldsymbol{\alpha}, t_k, \boldsymbol{x}))$, which complicates the solution of the integral. Researchers often use some smoothing functions such as Gaussian functions, Sinc functions, etc. to approximate the Dirac function [61]. Here, the Gaussian function is used, i.e.,

$$\delta(u_k - g(\boldsymbol{\alpha}, t_k, \boldsymbol{x})) \approx \phi(u_k - g(\boldsymbol{\alpha}, t_k, \boldsymbol{x}), \sigma) = \frac{1}{\sqrt{2\pi}\sigma} \exp\left(-\frac{(u_k - g(\boldsymbol{\alpha}, t_k, \boldsymbol{x}))^2}{2\sigma^2}\right), \tag{39}$$

where $\sigma$ is the smoothing parameter which can be determined according to Ref. [62]. As $\sigma$ approaches zero, there is $\lim_{\sigma \to 0} \phi(y, \sigma) = \delta(y)$. Using Eq. (39), the numerator in Eq. (27) can be expressed as

$$\int_{-\infty}^{+\infty} \begin{bmatrix} \rho_\alpha(\boldsymbol{\alpha}) \rho_{\beta_k}\left(z_k - \hat{\boldsymbol{h}}_k(\boldsymbol{\alpha}, \boldsymbol{x})\right) \\ \times \delta(u_k - g(\boldsymbol{\alpha}, t_k, \boldsymbol{x})) \end{bmatrix} d\boldsymbol{\alpha} \approx \int_{-\infty}^{+\infty} \begin{bmatrix} \rho_\alpha(\boldsymbol{\alpha}) \rho_{\beta_k}\left(z_k - \hat{\boldsymbol{h}}_k(\boldsymbol{\alpha}, \boldsymbol{x})\right) \\ \times \frac{1}{\sqrt{2\pi}\sigma} \exp\left(-\frac{(u_k - g(\boldsymbol{\alpha}, t_k, \boldsymbol{x}))^2}{2\sigma^2}\right) \end{bmatrix} d\boldsymbol{\alpha}. \tag{40}$$

Similarly, based on the known PDF $\rho_\alpha(\boldsymbol{\alpha})$ and the GQMC, we can generate $n$ samples $\boldsymbol{\alpha}_i$ and their non-equal weights $w_i$, $i = 1, 2, \cdots, n$. Substituting the samples into the stochastic dynamical equation (1) can give $n$ sets of $g(\boldsymbol{\alpha}_i, t_k, \boldsymbol{x})$ and $\hat{\boldsymbol{h}}_k(\boldsymbol{\alpha}_i, \boldsymbol{x})$. Using $g(\boldsymbol{\alpha}_i, t_k, \boldsymbol{x})$, $\hat{\boldsymbol{h}}_k(\boldsymbol{\alpha}_i, \boldsymbol{x})$, and Eq. (39), Eq. (40) can be approximated as

$$\int_{-\infty}^{+\infty} \begin{bmatrix} \rho_\alpha(\boldsymbol{\alpha}) \rho_{\beta_k}(z_k - \hat{\boldsymbol{h}}_k(\boldsymbol{\alpha}, \boldsymbol{x})) \\ \times \delta(u_k - g(\boldsymbol{\alpha}, t_k, \boldsymbol{x})) \end{bmatrix} d\boldsymbol{\alpha} \approx \sum_{i=1}^{n} w_i \begin{bmatrix} \rho_{\beta_k}(z_k - \hat{\boldsymbol{h}}_k(\boldsymbol{\alpha}_i, \boldsymbol{x})) \\ \times \frac{1}{\sqrt{2\pi}\sigma} \exp\left(-\frac{(u_k - g(\boldsymbol{\alpha}_i, t_k))^2}{2\sigma^2}\right) \end{bmatrix}. \quad (41)$$

The substitution of Eqs. (31) and (41) into Eq. (27) gives:

$$\rho_{u_k|z_k}(u_k | z_k) = \frac{\sum_{i=1}^{n} w_i \left[ \rho_{\beta_k}(z_k - \hat{\boldsymbol{h}}_k(\boldsymbol{\alpha}_i, \boldsymbol{x})) \frac{1}{\sqrt{2\pi}\sigma} \exp\left(-\frac{(u_k - g(\boldsymbol{\alpha}_i, t_k, \boldsymbol{x}))^2}{2\sigma^2}\right) \right]}{\sum_{i=1}^{n} w_i \rho_{\beta_k}(z_k - \hat{\boldsymbol{h}}_k(\boldsymbol{\alpha}_i, \boldsymbol{x}))}. \quad (42)$$

In terms of Eq. (42), the KCQ-PDF of the random response $u_k$ considering key conditions can be calculated.

### 4.3 Offline-online coupled computational strategy

As shown in Eqs. (34), (35) and (42), the most computationally intensive part in the calculation of KCQs lies in calculating the dynamical response values $g(\boldsymbol{\alpha}_i, t_k)$ for all samples $\boldsymbol{\alpha}_i$. It is necessary to solve the dynamical equation (1) $n$ times, which is a very large amount of calculation, especially when involving nonlinear structural dynamical problems. In fact, after all samples $\boldsymbol{\alpha}_i$ are generated, their corresponding responses $g(\boldsymbol{\alpha}_i, t_k)$ are not affected by the measurement data and can be calculated directly in advance. Therefore, in order to improve the computational efficiency of KCQ, this subsection proposes an offline-online coupled computing method which includes two stages: offline database generation and online conditional uncertainty quantification. Offline database generation stage provides a database of uncertain responses corresponding to samples, and its calculation steps are as follows:

(1) According to the PDF $\rho_\alpha(\alpha)$ and the GQMC method, generate $n$ samples $\alpha_i$, $i = 1, 2, \cdots, n$, and their corresponding non-equal weights $w_i$;

(2) Substitute each sample $\alpha_i$ into the dynamical equation (1) to get $n$ solutions of the dynamical response values $g(\alpha_i, t_k)$ at time $t_k$, $k = 0, 2, \cdots, N_T$;

(3) Using Eq. (8) to further obtain $h(g(\alpha_i, t_k))$.

During the offline database generation stage, a database containing $\{w_i, \alpha_i, g(\alpha_i, t_k), h(g(\alpha_i, t_k))\}_{\substack{i=1,\cdots,n \\ k=0,\cdots,N_T}}$ can be obtained. Next, the online conditional uncertainty quantification stage can be carried out:

(1) Based on the offline database $\{w_i, \alpha_i, g(\alpha_i, t_k), h(g(\alpha_i, t_k))\}_{\substack{i=1,\cdots,n \\ k=0,\cdots,N_T}}$ and the measurement data $y_{1:k}$, calculate the key matrix $p_k(x)$ at position $x$ at time $t_k$ through the correlation coefficient shown in Eq. (19);

(2) Calculate $z_k(x) = p_k(x) y_{1:k}$, $\hat{h}_k(\alpha, x)$, and $\rho_{\beta_k}(z_k - \hat{h}_k(\alpha_i, x))$ in terms of Eq. (22);

(3) Based on Eqs. (34), (35) and (42), calculate the KCQ-mean $\hat{u}_k(x)$, the KCQ-variance $P_k(x)$ and the KCQ-PDF $\rho_{u_k|z_k}(u_k | z_k)$ of the stochastic response $u_k(x)$.

## 5. Numerical examples

### 5.1 Mass-Spring-Damper System

This example considers a single-degree-of-freedom mass-spring-damper system, whose equation of free vibration is:

$$M\ddot{u} + C(\varepsilon_1)\dot{u} + K(\varepsilon_2)u = f(t), \tag{43}$$

where the mass $M = 5 \text{ kg}$, damping $C = 5(1+\varepsilon_1) \text{ N}/(\text{m/s})$, stiffness $K = 11(1+\varepsilon_2) \text{ N/m}$, and restoring force $f(t) = 10\sin(3t) \text{ N}$. $\boldsymbol{\varepsilon} = [\varepsilon_1, \varepsilon_2]^\text{T}$ are independent random variables that follow a normal distribution with a mean of $\boldsymbol{\mu}_\varepsilon = (0, 0) \text{ m}$ and a standard deviation $\boldsymbol{\sigma}_\varepsilon = (0.2, 0.2) \text{ m}$. The initial displacement is $u_0 = 0 \text{ m}$ and the initial velocity is $\dot{u}_0 = 0 \text{ m/s}$. In this example, due to the slight uncertainty in the damping $C$ and stiffness $K$, the structural response also exhibits uncertainty, which necessitates the estimation of the statistical properties of the response. In this example, the velocity of system $\dot{u}(t_k)$ is measured, and the measurement data is denoted as $y(t_k)$, as shown in Fig. 1. It is assumed that the measurement data contains errors, denoted as $y(t_k) = \dot{u}(t_k) + v(t_k)$, where $v(t_k)$ represents the measurement error. This error is a Gaussian noise with a mean of $0 \text{ m/s}$ and a standard deviation (SD) $0.03 \text{ m/s}$.

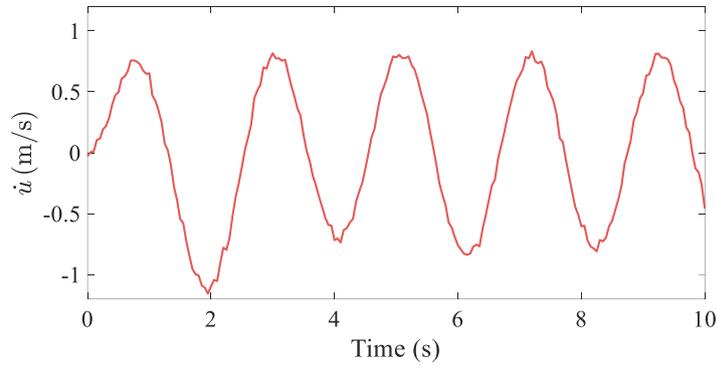

Fig. 1. Velocity measurement data.

To verify the correctness of the KCQ numerical schemes (42), (34), and (35) proposed in Section 4, this example will use the GQMC method proposed by Chen and Li (KCQ-CL), and the GQMC method proposed by Wu and Zhao (KCQ-WZ) to calculate Eqs. (42), (34), and (35), respectively. The symplectic Euler midpoint scheme is used for time integration. The time length for the dynamic simulation is $T = 10 \text{ s}$, the time step is $\Delta t = 0.05 \text{ s}$. Both KCQ-WZ and KCQ-CL use 500 samples for calculation. For comparison, this example also employs the MC method with 1,000,000 samples to compute the high-dimensional integrals in Eqs. (27)-(29) and this method is denoted as KCQ-MC. The computational results of KCQ-MC are taken as the reference solution. The key condition numbers for KCQ-WZ, KCQ-CL, and KCQ-MC are all taken as $N_k = 2$. To compare the conditional uncertainty quantification method KCQ proposed in this paper with the traditional non-conditional uncertainty quantification, this example also uses the traditional non-conditional MC method (called N-MC) with 1,000,000 samples to directly assess the statistical properties of the random response.

First, the accuracy of the numerical schemes proposed in Subsection 4.1 is verified. Table 1 provides the deflection values calculated by KCQ-WZ and KCQ-CL at $t = 2.5, 5, 7.5, 10 \text{ s}$, the reference solutions calculated by KCQ-MC, the relative errors (RE, see data in parentheses) compared to the KCQ-MC method, as well as the CPU times for different methods. CPU times are the sum of offline computation times and online computation times, with the offline computation times including the point set generation times and the dynamic response computation times. The KCQ-PDF curves for displacement and velocity at time $t = 2.5, 5, 7.5, 10 \text{ s}$ calculated using the KCQ-WZ, KCQ-CL, and KCQ-MC methods are plotted in Figs. 2 and 3, respectively.

From Table 1, it can be seen that compared to the reference solution, using KCQ-WZ and KCQ-CL with only 500 sample points, the results at each iteration step are highly accurate, with REs all less than 5%. The calculation results of CPU times show that for the response calculation problem of this mass-spring-damper system, both GQMC-based KCQ-WZ and KCQ-CL are significantly faster than the KCQ-MC method. KCQ-WZ can speed up the calculations by 3 orders of magnitude, and KCQ-CL by 2 orders of magnitude. In addition, the computational results indicate that the offline database computation steps of the two KCQs can indeed share the computational load of the entire process, thereby reducing the computational burden of the online computation steps. Figs. 2 and 3 show that the KCQ-PDF curves for displacement and velocity at each iteration step calculated based on KCQ-WZ and KCQ-CL can well match the KCQ-PDF curves for displacement and velocity calculated based on KCQ-MC, demonstrating the correctness of the proposed KCQ numerical schemes.

**Table 1**

Comparison of the computational results of KCQ-WZ and KCQ-CL with the reference results.

| Index | Time | $u(t)$ (m) | | | $\dot{u}(t)$ (m/s) | | |
|---|---|---|---|---|---|---|---|
| | | KCQ-MC | KCQ-WZ | KCQ-CL | KCQ-MC | KCQ-WZ | KCQ-CL |
| KCQ-mean | $t=2.5$ s | -0.3396 | -0.3399 (0.08%) | -0.3391 (0.13%) | -0.1106 | -0.1109 (0.26%) | -0.1103 (0.29%) |
| | $t=5$ s | -0.0479 | -0.0478 (0.35%) | -0.0481 (0.35%) | 0.7965 | 0.7965 (0.01%) | 0.7965 (0.00%) |
| | $t=7.5$ s | 0.2016 | 0.2015 (0.03%) | 0.2017 (0.05%) | 0.4823 | 0.4824 (0.02%) | 0.4822 (0.02%) |
| | $t=10$ s | 0.2182 | 0.2181 (0.04%) | 0.2182 (0.03%) | -0.4127 | -0.4130 (0.09%) | -0.4126 (0.02%) |
| KCQ-SD | $t=2.5$ s | 0.0145 | 0.0144 (0.48%) | 0.0148 (2.47%) | 0.0204 | 0.0201 (1.73%) | 0.0209 (2.50%) |
| | $t=5$ s | 0.0202 | 0.0202 (0.06%) | 0.0203 (0.64%) | 0.0131 | 0.0132 (0.78%) | 0.0132 (0.74%) |
| | $t=7.5$ s | 0.0049 | 0.0051 (4.26%) | 0.0049 (0.58%) | 0.0181 | 0.0180 (0.31%) | 0.0178 (1.26%) |
| | $t=10$ s | 0.0065 | 0.0066 (0.34%) | 0.0065 (0.57%) | 0.0176 | 0.0176 (0.08%) | 0.0172 (2.28%) |
| CPU Times (s) | | 652.01+49.119 | 0.25+0.025 | 7.44+0.022 | 652.01+48.800 | 0.25+0.021 | 7.44+0.023 |

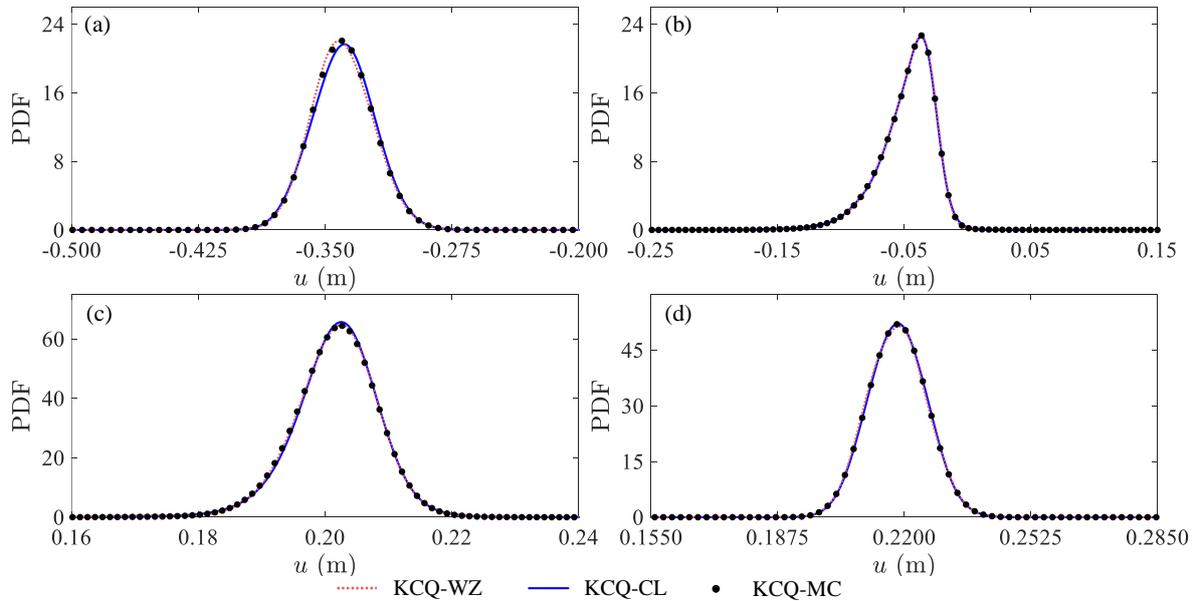

Fig. 2. KCQ-PDF curves of displacement at different times: (a) $t=2.5$ s; (b) $t=5$ s; (c) $t=7.5$ s; (d) $t=10$ s.

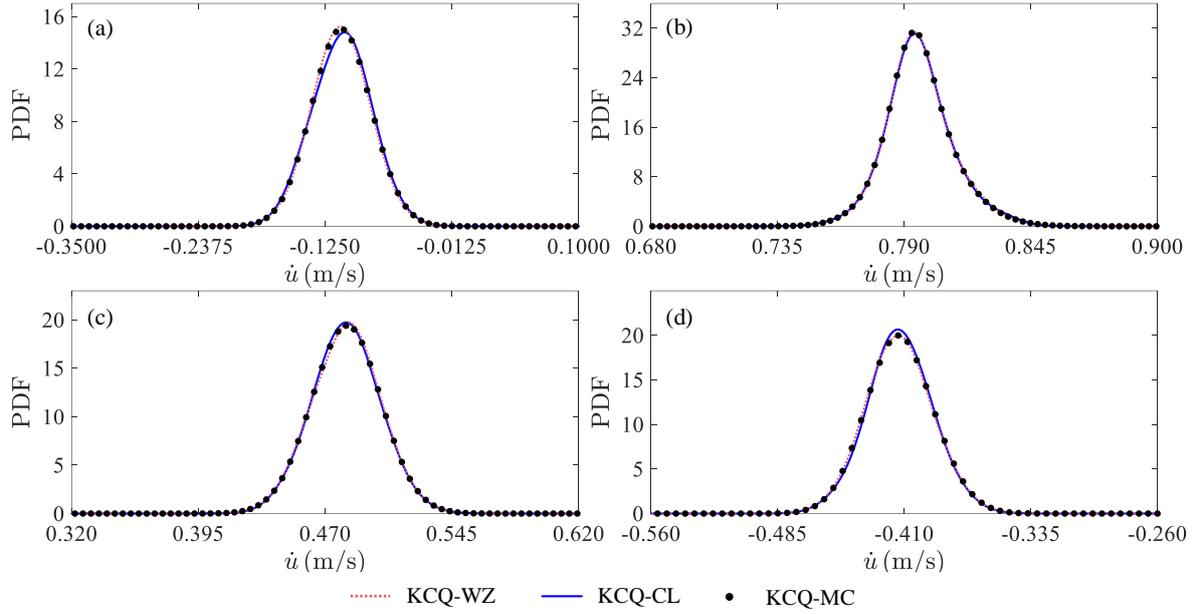

Fig. 3. KCQ-PDF curves of velocity at different times: (a) $t = 2.5$ s; (b) $t = 5$ s; (c) $t = 7.5$ s; (d) $t = 10$ s.

The most significant different between KCQ and traditional quantification method is whether to consider the measurement data conditions. To clarify the differences between the KCQ method and traditional uncertainty quantification methods, we compare the KCQ-means, KCQ-SDs, and KCQ-PDFs calculated by KCQ-WZ calculation with the means, SDs, and PDFs calculated by the traditional N-MC without considering measurement conditions. The means of displacement and velocity calculated by the N-MC method without considering measurement conditions, as well as the three-standard deviation interval (TSDI, i.e., $\text{mean}_{\text{N-MC}} \pm 3\text{SD}_{\text{N-MC}}$), are plotted in Fig. 4. For the convenience of comparison, we also plot the KCQ means of displacement and velocity calculated by the proposed KCQ-WZ, as well as the corresponding KCQ-TSDI, in Fig. 4. In addition, we also plotted the PDFs of displacement and velocity of the responses calculated by KCQ-WZ and N-MC at $t = 2.5, 5, 7.5, 10$ s in Figs. 5 and 6, respectively. It is very evident from Figs. 4-6 that the KCQ-TSDI ($\text{mean}_{\text{KCQ-WZ}} \pm 3\text{SD}_{\text{KCQ-WZ}}$) and KCQ-PDFs at different times obtained from the KCQ-WZ method, which takes measurement conditions into account, are significantly narrower than the results from the traditional N-MC method. This indicates that for this problem, considering the measurement conditions has led to a more refined quantification of the uncertainty in the random structural response, with smaller SD and narrower probability distribution intervals. This means that the KCQ method can reduce the level of uncertainty in the response by considering these conditions, providing more precise and reliable statistical results.

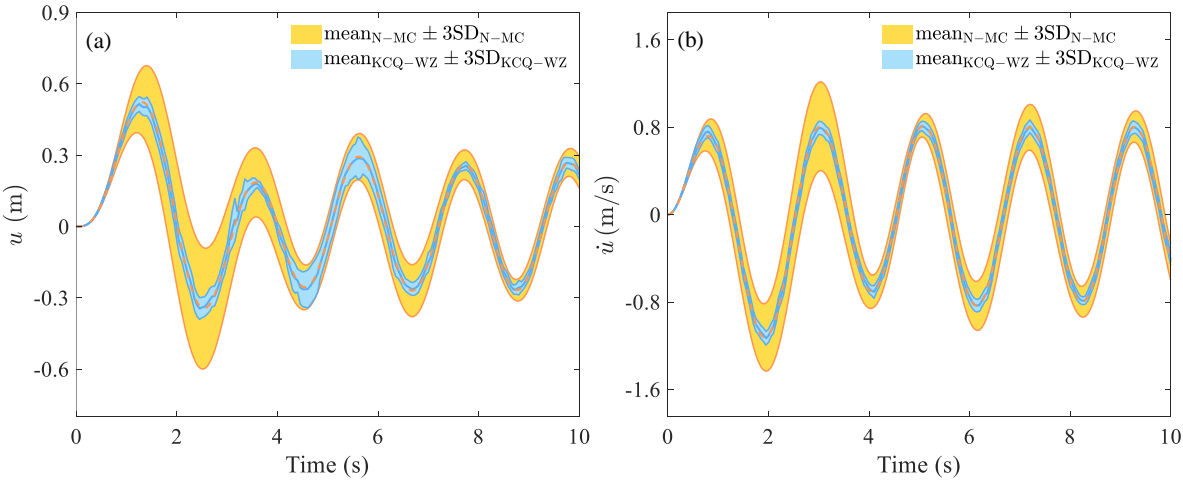

Fig. 4. TSDIs of KCQ-WZ and N-MC: (a) Displacement; (b) Velocity.

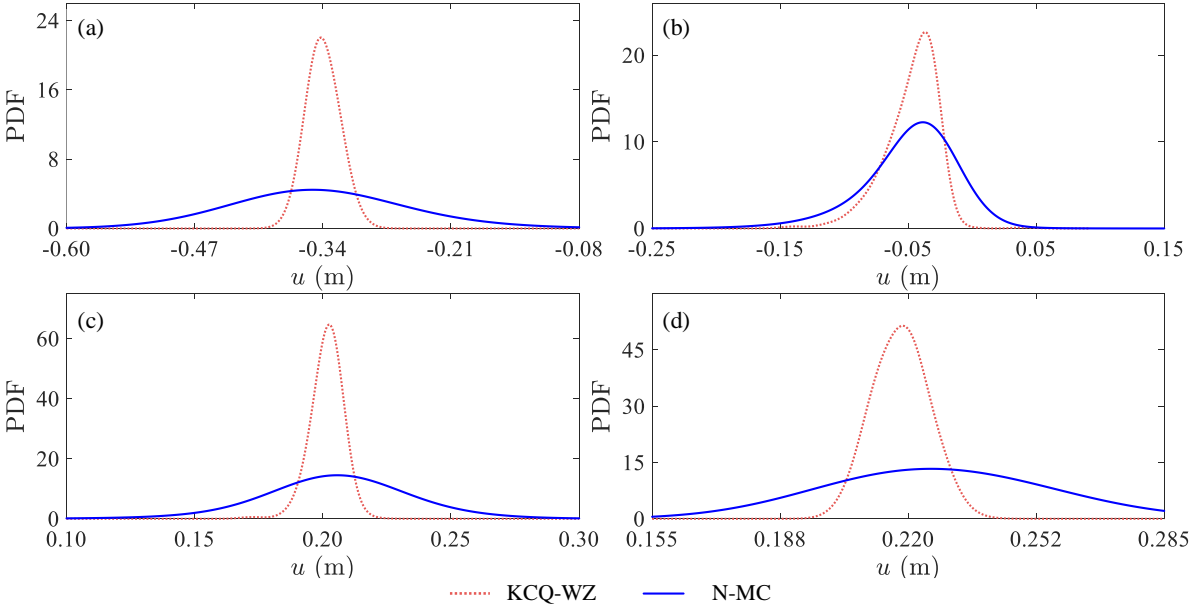

Fig. 5. PDF curves of displacement at different times: (a) $t = 2.5$ s; (b) $t = 5$ s; (c) $t = 7.5$ s; (d) $t = 10$ s.

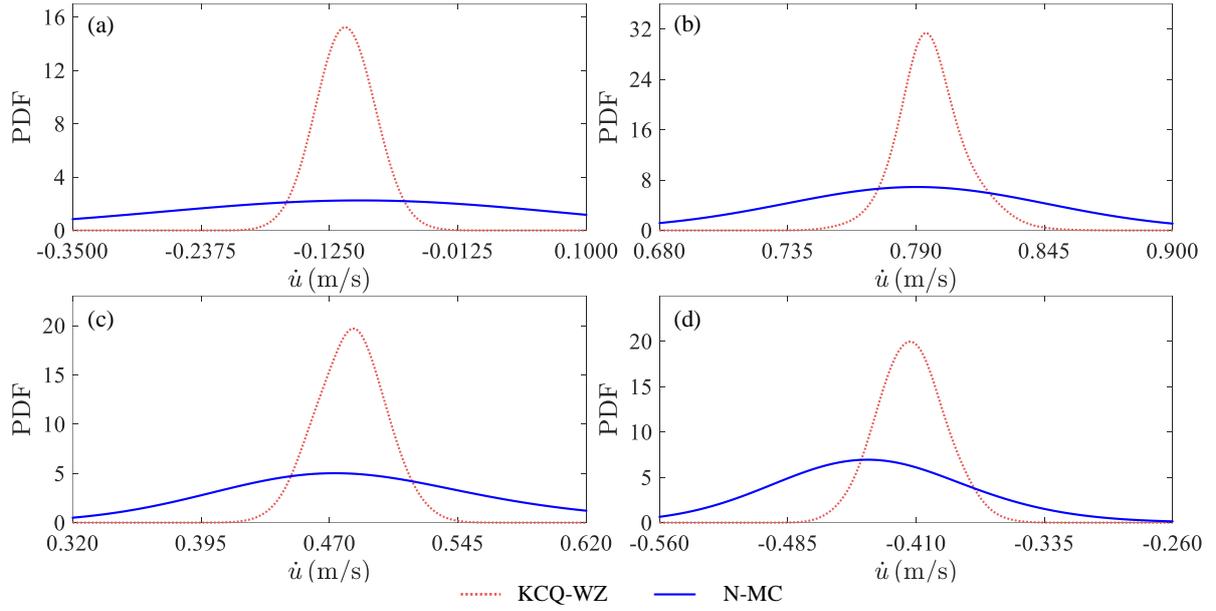

Fig. 6. PDF curves of velocity at different times: (a) $t = 2.5\,\text{s}$; (b) $t = 5\,\text{s}$; (c) $t = 7.5\,\text{s}$; (d) $t = 10\,\text{s}$.

**5.2 A geometrically nonlinear beam**

The second example involves the dynamical problem of a geometrically nonlinear cantilever beam, the model of which is shown in Fig. 7. The cross-sectional size of the beam is $0.1\,\text{m} \times 0.1\,\text{m}$, the linear density of the beam is $d = 100\,\text{kg/m}$, and the length of the beam is $L = 3\,\text{m}$. The beam is fixed at the left end and free at the right end. The entire beam is subjected to a deterministic uniformly distributed load of $q = 5 \times 10^4\,\text{N/m}$.

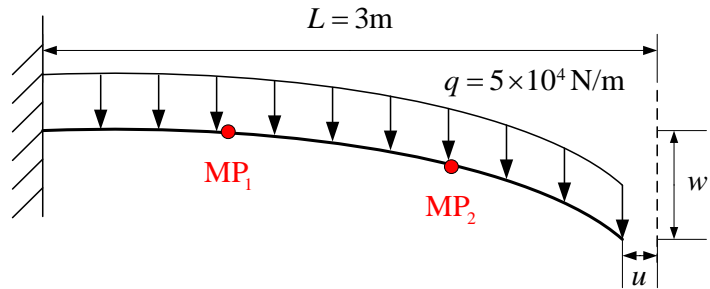

Fig. 7. Geometrically nonlinear cantilever beam subjected to the uniformly distributed load.

According to the elastic mechanics [63], The vibration equation for this beam is:

$$\begin{cases} d\ddot{u} = c_u \dot{u} + \dfrac{\partial(EAu_x)}{\partial x} + \dfrac{1}{2}\dfrac{\partial(EAw_x^2)}{\partial x} \\ d\ddot{w} + \dfrac{\partial^2(EIw_{xx})}{\partial x^2} = c_w \dot{w} + \dfrac{1}{2}\dfrac{\partial(EAw_x^3)}{\partial x} + \dfrac{\partial(EAu_x w_x)}{\partial x} + q \end{cases}, \tag{44}$$

where, $u$ and $w$ are the transverse and vertical displacements of the beam. $E$, $A$ and $I$ are the elastic modulus, cross-sectional area, and moment of inertia of the beam, respectively. $c_u$ and $c_w$ are the damping coefficients, they are set to $c_u = c_w = 40d\ \text{N}/(\text{m/s})$. Assume that the elastic modulus $E$ is a stationary random field, and can be represented by the K-L expansion:

$$E(x,\varepsilon) = E_0 + \sum_{i=1}^{M} \varepsilon_i \sqrt{\kappa_i} f_i(x), \tag{45}$$

in which $E_0 = 2\times 10^{11}\,\text{Pa}$ is the mean of $E$, $\kappa_i$ is the $i$-th eigenvalue of the covariance function to the random field, and $f_i(x)$ is the eigenfunction corresponding to $\kappa_i$. $\varepsilon_i$ are zero-mean random variables, independent of each other, and follow a standard normal distribution. $M$ is the order of the K-L expansion, and in this example $M = 10$. According to Ref. [64], $\kappa_i$ and $f_i(x)$ can be calculated using the following formula:

$$\kappa_i = \dfrac{2c_K \sigma_E^2}{\omega_i^2 + c_K^2}, \quad f_i(x) = \begin{cases} \dfrac{\cos(\omega_i x)}{\sqrt{a_K + \dfrac{\sin(2\omega_i a_K)}{2\omega_i}}}, & i \text{ is odd} \\ \dfrac{\sin(\omega_i x)}{\sqrt{a_K - \dfrac{\sin(2\omega_i a_K)}{2\omega_i}}}, & i \text{ is even} \end{cases}, \tag{46}$$

if $i$ is odd, $\omega_i \tan(a_K \omega_i) = c_K$; if $i$ is even, $\omega_i = -c_K \tan(a_K \omega_i)$, $a_K = 3$, $c_K = 0.333$, and $\sigma_E = 0.2$. The beam is uniformly divided into ten elements. The finite element equation is used to spatially discretize equation (44), and the symplectic Euler midpoint scheme is used for time integration. The time length for the dynamic simulation is $T = 0.4\,\text{s}$, the time step is $\Delta t = 0.001\,\text{s}$. Two measurement points are set up at positions $x = 0.9\,\text{m}$ and $x = 2.1\,\text{m}$ (as shown by the red points in Fig. 7), and the measured deflections are obtained as $y(t_k)$, as shown in Fig. 8. It is also assumed that the measured deflections contain measurement errors $y(t_k) = w(t_k) + v(t_k)$, where the noise $v(t_k)$ is an independent Gaussian process with a mean of 0 m and a SD of 0.005 m.

KCQ-mean, KCQ-SD, and KCQ-PDF for the random vertical displacement and velocity at the end of the beam are calculated using KCQ-WZ, KCQ-CL, and KCQ-MC, respectively, with KCQ-WZ and KCQ-CL using 600 samples for the calculation. KCQ-MC uses 100,000 samples, and its results are taken as the reference solution. The number of key conditions used in all three methods is set to $N_k = 1$. Additionally, to compare the proposed KCQ method with traditional non-conditional method, this example also employs the traditional N-MC based on 100,000 samples to directly assess the mean, SD, and PDF of the random deflection of the beam.

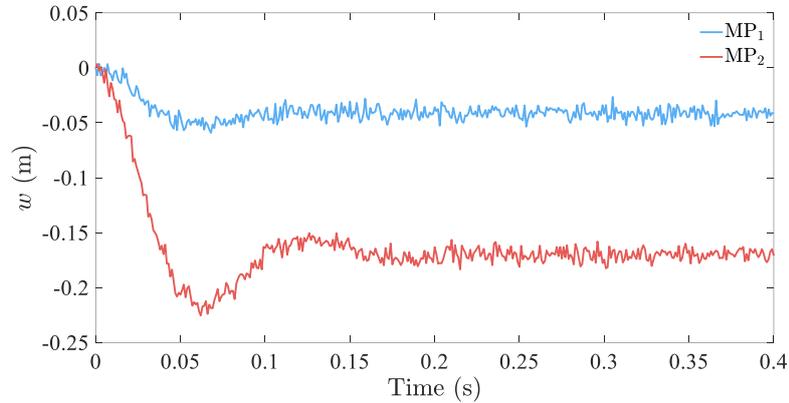

Fig. 8. Deflection data measured at the two measurement points.

Table 2 provides the deflections at the end of beam calculated by KCQ-WZ and KCQ-CL at time $t = 0.1, 0.2, 0.3, 0.4 \text{ s}$, the RE, as well as the CPU times. The KCQ-PDF curves for vertical deflection and velocity at time $t = 0.1, 0.2, 0.3, 0.4 \text{ s}$ calculated using the KCQ-WZ, KCQ-CL, and KCQ-MC methods are plotted in Figs. 9 and 10, respectively.

**Table 2**

Comparison of the computational results of KCQ-WZ and KCQ-CL with the reference results.

| Index | Time | $w(L)$ | | | $\dot{w}(L)$ | | |
|---|---|---|---|---|---|---|---|
| | | KCQ-MC | KCQ-WZ | KCQ-CL | KCQ-MC | KCQ-WZ | KCQ-CL |
| KCQ-mean | $t=0.1$ s | -0.2927 | -0.2927 (0.00%) | -0.2923 (0.11%) | 2.1803 | 2.1689 (0.52%) | 2.2010 (0.95%) |
| | $t=0.2$ s | -0.2707 | -0.2707 (0.02%) | -0.2715 (0.31%) | 0.2413 | 0.2407 (0.27%) | 0.2399 (0.57%) |
| | $t=0.3$ s | -0.2879 | -0.2880 (0.02%) | -0.2874 (0.19%) | -0.0220 | -0.0222 (0.86%) | -0.0219 (0.25%) |
| | $t=0.4$ s | -0.2920 | -0.2921 (0.06%) | -0.2910 (0.34%) | -0.0049 | -0.0049 (0.25%) | -0.0050 (1.90%) |
| KCQ-SD | $t=0.1$ s | 0.0092 | 0.0092 (0.88%) | 0.0088 (4.38%) | 0.1314 | 0.1326 (0.91%) | 0.1304 (0.78%) |
| | $t=0.2$ s | 0.0086 | 0.0083 (3.40%) | 0.0089 (3.56%) | 0.0258 | 0.0254 (1.72%) | 0.0259 (0.33%) |
| | $t=0.3$ s | 0.0091 | 0.0091 (0.95%) | 0.0085 (5.76%) | 0.0083 | 0.0085 (2.97%) | 0.0080 (3.18%) |
| | $t=0.4$ s | 0.0092 | 0.0092 (0.30%) | 0.0087 (4.90%) | 0.0007 | 0.0006 (1.27%) | 0.0006 (4.90%) |
| CPU Times (s) | | 289293.8+101.6 | 1773.6+1.5 | 1838.2+1.8 | 289293.8+101.2 | 1773.6+1.5 | 1838.2+1.7 |

It can be seen from Table 2 that compared to the reference solution, the KCQ-mean and KCQ-SD calculated at each iteration step using KCQ-WZ and KCQ-CL both have very high accuracy, with REs all less than 5%. At the same time, the CPU Times indicate that for this dynamic response uncertainty quantification problem of the beam considering geometric nonlinearity, both GQMC-based KCQ-WZ and KCQ-CL are significantly faster than the KCQ-MC, with KCQ-WZ and KCQ-CL being two orders of magnitude faster. Additionally, the offline database calculation steps for both KCQ methods account for about 99.99% of the entire computation process, thus the offline-online coupled computational strategy proposed in Section 4 is indeed effective in reducing the computational burden of the online calculation steps. From Figs. 9 and 10, it can be seen that the PDF curves for displacement and velocity at each iteration step calculated based on KCQ-WZ and KCQ-CL match well with the displacement and velocity PDF curves obtained from KCQ-MC calculations, demonstrating the correctness of the numerical schemes (42). Combining the results from Table 2 and Figs. 9 and 10, it can be concluded that the GQMC sampling and offline-online coupled computational strategy can effectively calculate the high-dimensional numerical integration in KCQ, providing great convenience for the computation of conditional mean, conditional SD, and conditional PDF of the response considering measurement conditions.

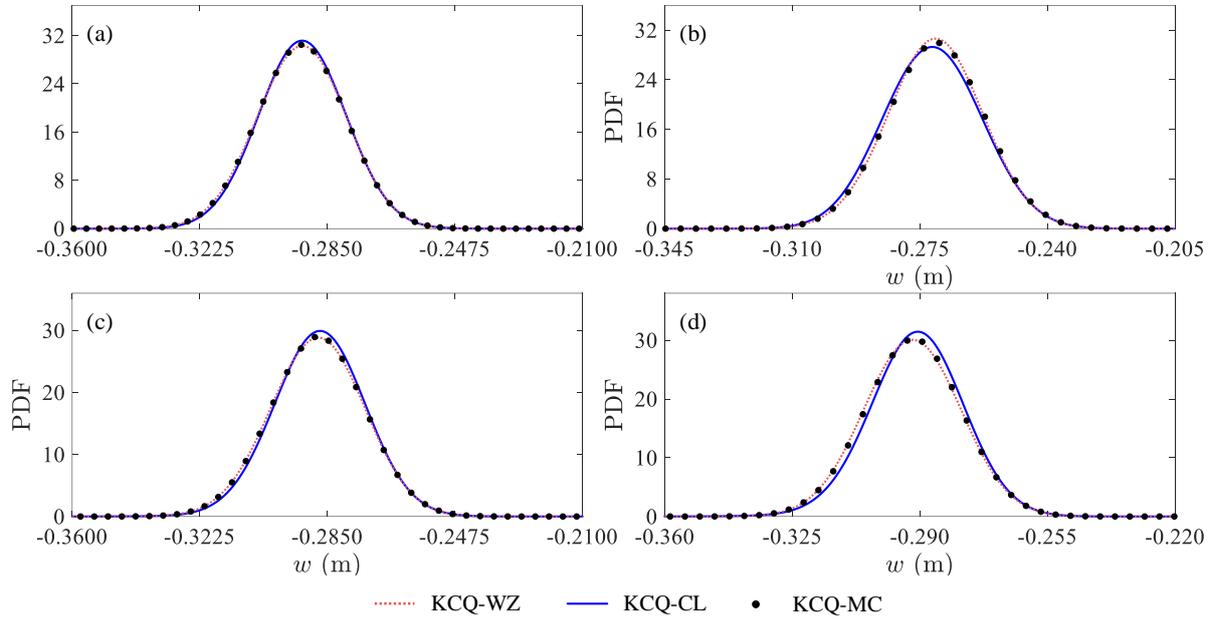

Fig. 9. KCQ-PDF curves of vertical displacement at the end of the beam at different times: (a) $t = 0.1\,\text{s}$; (b) $t = 0.2\,\text{s}$; (c) $t = 0.3\,\text{s}$; (d) $t = 0.4\,\text{s}$.

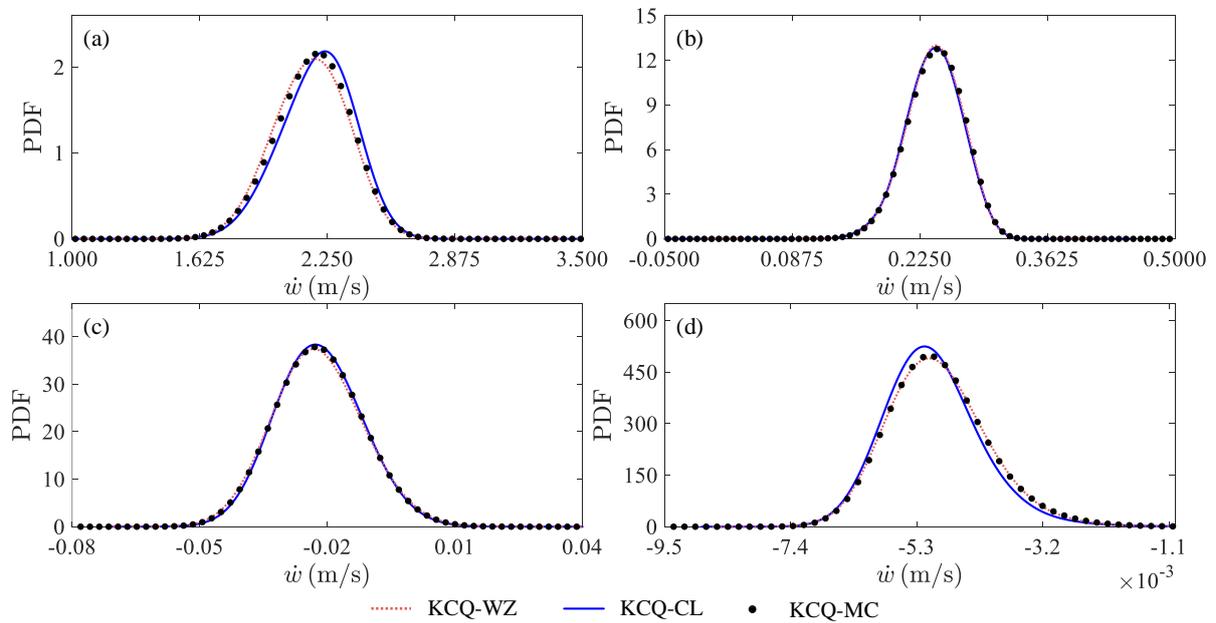

Fig. 10. KCQ-PDF curves of vertical velocity at the end of the beam at different times: (a) $t = 0.1\,\text{s}$; (b) $t = 0.2\,\text{s}$; (c) $t = 0.3\,\text{s}$; (d) $t = 0.4\,\text{s}$.

Next, the mean, SD, and PDF of the random response were calculated using both the KCQ-

WZ method and the traditional N-MC method without considering measurement conditions. First, the vertical displacement and velocity of the beam affected by the uncertainty of the elastic modulus were compared at $t = 0.4$ s using the KCQ-WZ and N-MC methods, as shown in Fig. 11. It is very evident from Fig. 11 that both the vertical displacement and velocity have much narrower TSDIs when calculated using the KCQ-WZ method, which includes measurement conditions, compared to the traditional N-MC method that does not consider measurement conditions. This indicates that for this problem, the responses calculated using the conditional method are more tightly clustered around their mean, with a smaller SD and a more precise distribution interval. We further compared the TSDIs of the vertical displacement and velocity at the end of the beam at $x = 3$ m at various times and plotted the results in Fig. 12. It is clear that Fig. 12 follows the same pattern as Fig. 11, with the TSDIs calculated using the KCQ-WZ method, which includes measurement conditions, being much narrower than those calculated using the traditional N-MC method that does not consider measurement conditions. The PDFs of the vertical displacement and velocity at the end of the beam at $t = 0.1, 0.2, 0.3, 0.4$ s calculated using KCQ-WZ and N-MC are plotted in Figs. 13 and 14, respectively. It is very evident from Figs. 13 and 14 that the PDF curves calculated using the KCQ-WZ method, which includes measurement conditions, are significantly narrower than those calculated using the traditional N-MC method that does not consider measurement conditions. In summary, for this problem, the responses calculated by considering measurement conditions are more tightly clustered around their mean, with a smaller SD. This means that the proposed KCQ method proposed can reduce the uncertainty of the response by considering measurement conditions, providing more precise and reliable statistics.

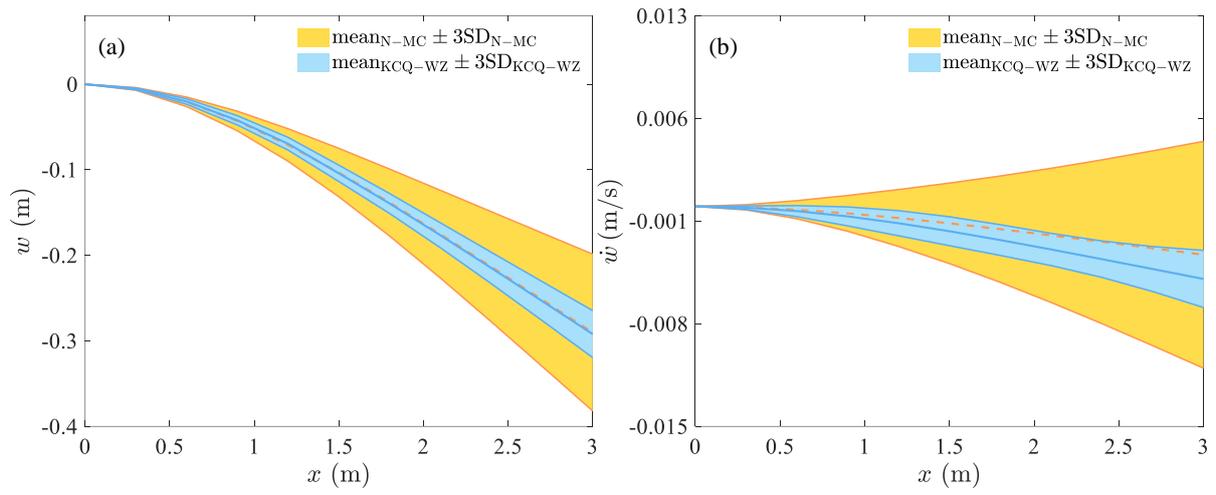

Fig. 11. TSDIs of KCQ-WZ and N-MC at $t = 0.4$ s : (a) Vertical displacement; (b) Vertical velocity.

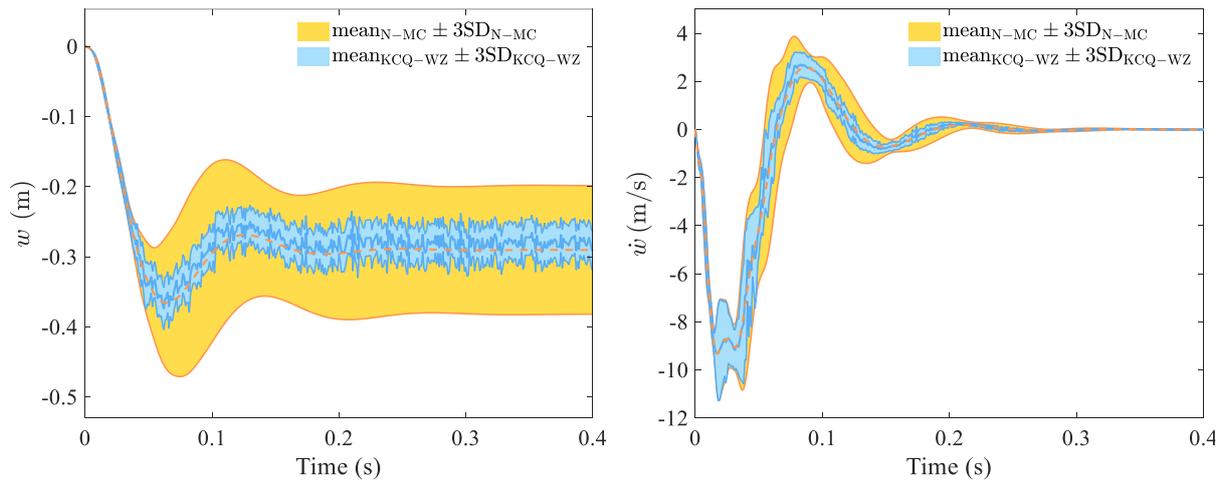

Fig. 12. TSDIs of KCQ-WZ and N-MC at $x = 3$ m : (a) Vertical displacement; (b) Vertical velocity.

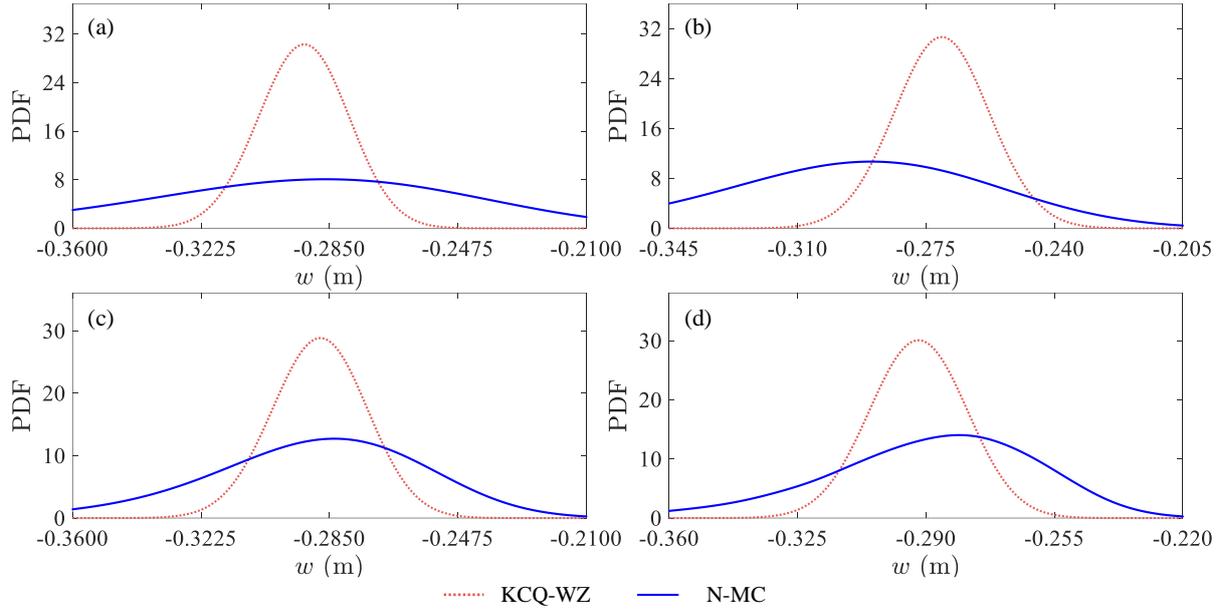

Fig. 13. PDF curves of the vertical displacement at the end of the beam at different times: (a) $t = 0.1\,\text{s}$; (b) $t = 0.2\,\text{s}$; (c) $t = 0.3\,\text{s}$; (d) $t = 0.4\,\text{s}$.

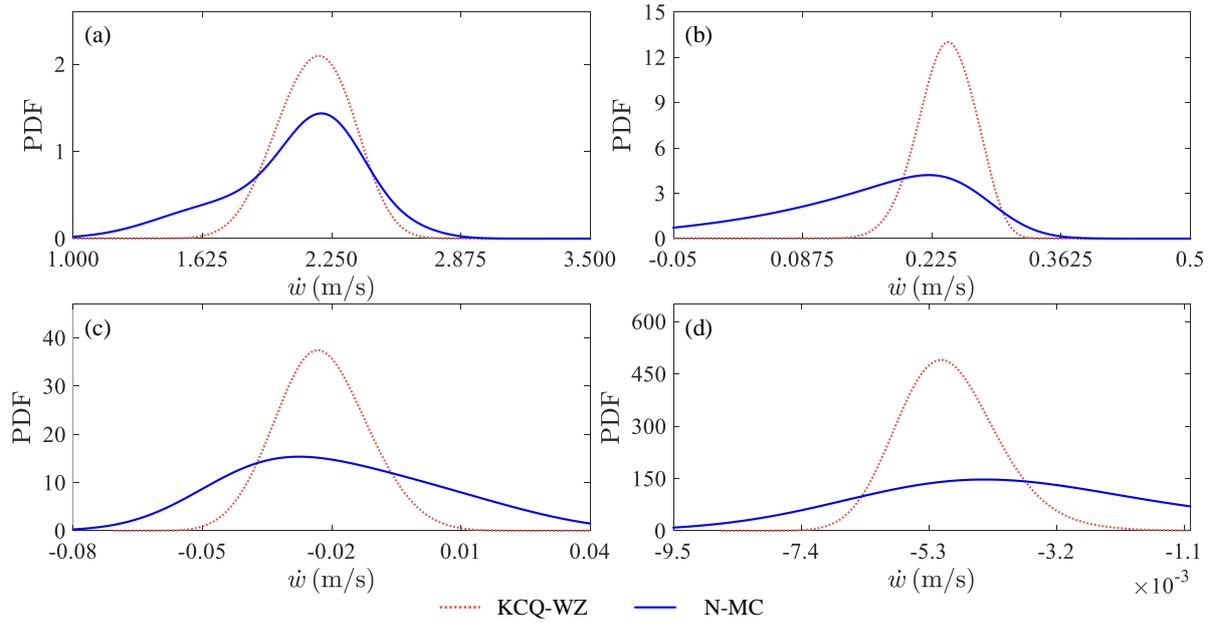

Fig. 14. PDF curves of the vertical velocity at the end of the beam: (a) $t = 0.1\,\text{s}$; (b) $t = 0.2\,\text{s}$; (c) $t = 0.3\,\text{s}$; (d) $t = 0.4\,\text{s}$.

## 5.3 Concrete bridge

Fig. 15. Concrete bridge model.

In this section, a finite element model of a concrete bridge is considered, as shown in Fig. 15. The steel frame of the bridge consists of 38 6-degree-of-freedom beam elements with an elastic modulus of $E_b = 2.06 \times 10^{11} \text{ N/m}^2$, a shear modulus of $G_b = 8.11 \times 10^{10} \text{ N/m}^2$, and a density of $\rho_b = 7850 \text{ kg/m}^3$, with a square cross-section of side length 1m, and the shear coefficient is not considered. The concrete bridge deck contains 224 6-degree-of-freedom shell elements with a thickness of 0.5m, a Poisson's ratio of 0.25, and a density of $\rho_s = 3600 \text{ kg/m}^3$. It is assumed that the elastic modulus of the concrete bridge deck is a stationary normal random field, which can be expressed as the following K-L expansion:

$$E_s(\mathbf{x}, \boldsymbol{\xi}) = E_{s,0} + \sigma_{E_s} \sum_{j=1}^{N_v} \xi_j \sqrt{\tilde{\lambda}_j} \tilde{\phi}_j(\mathbf{x}) = E_{s,0} + \sum_{j=1}^{N_v} E_{s,i}(\mathbf{x}) \xi_j, \sigma_{E_s} = E_{s,0} V_c, \quad (47)$$

where $E_{s,0} = 3.45 \times 10^{10} \text{ N/m}^2$ represents the mean value of the elastic modulus for the shell elements, the coefficient of variation is $V_c = 0.2$, and the covariance function is defined as D.

$$C(\mathbf{x}_i, \mathbf{x}_j) = \exp\left(-\frac{|x_i - x_j|_1}{20} - \frac{|y_i - y_j|}{140}\right). \quad (48)$$

A uniformly distributed load that varies with time is applied to the bridge deck:

$$P(t) = -1.0 \times 10^5 \left[ \sin\left(0.02(20-t)^2\right) - \sin\left(0.02(20+t)^2\right) \right] \text{N/m}^2. \tag{49}$$

The Rayleigh damping is adopted with a damping ratio of $\zeta = 0.03$. The time step is taken as $\Delta t = 0.02$ s, and the total time integration is taken as $T = 20$ s. Three measurement points $C_1$, $C_2$, and $C_3$ are arranged on the concrete bridge (as shown by the red points in Fig. 15). The measured vertical displacement is $y(t_k) = w(t_k) + v(t_k)$, as shown in Fig. 16. The $v(t_k)$ follows a Gaussian distribution with a mean of 0 m and a SD of 0.003 m.

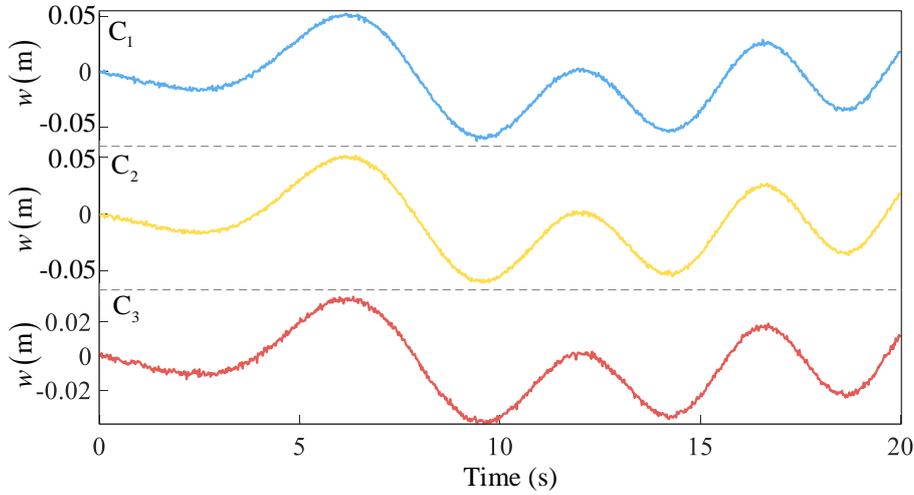

Fig. 16. Vertical displacement data measured at three measurement points: $C_1, C_2$ and $C_3$.

The KCQ-mean, KCQ-SD, and KCQ-PDF for the random vertical displacement and vertical velocity at points A and B on the concrete bridge are calculated using KCQ-WZ, KCQ-CL, and KCQ-MC, respectively, with KCQ-WZ and KCQ-CL using 500 samples for the calculation. KCQ-MC uses 100,000 samples, and its results are taken as the reference solution. The number of key conditions used in all three methods is set to the same value. Additionally, to compare the proposed KCQ method with the traditional non-conditional uncertainty quantification, this example also employs the traditional N-MC method based on 100,000 samples to directly assess the degree of uncertainty in the vertical displacement and velocity at points A and B on the concrete bridge.

**Table 3**

Comparison of the computational results at point A of KCQ-WZ and KCQ-CL with the reference results.

| Index | Time | $w(A)$ | | | $\dot{w}(A)$ | | |
|---|---|---|---|---|---|---|---|
| | | KCQ-MC | KCQ-WZ | KCQ-CL | KCQ-MC | KCQ-WZ | KCQ-CL |
| KCQ-mean | $t=5\text{ s}$ | 0.0550 | 0.0549 (0.11%) | 0.0548 (0.41%) | 0.0516 | 0.0511 (0.90%) | 0.0514 (0.35%) |
| | $t=10\text{ s}$ | -0.1018 | -0.1017 (0.15%) | -0.1014 (0.47%) | 0.0383 | 0.0383 (0.10%) | 0.0382 (0.40%) |
| | $t=15\text{ s}$ | -0.0661 | -0.0660 (0.14%) | -0.0658 (0.39%) | 0.0800 | 0.0799 (0.14%) | 0.0797 (0.39%) |
| | $t=20\text{ s}$ | 0.0337 | 0.0342 (1.49%) | 0.0335 (0.50%) | 0.0834 | 0.0846 (1.51%) | 0.0829 (0.51%) |
| KCQ-SD | $t=5\text{ s}$ | 0.0035 | 0.0035 (0.38%) | 0.0035 (2.25%) | 0.0039 | 0.0038 (2.47%) | 0.0039 (2.33%) |
| | $t=10\text{ s}$ | 0.0048 | 0.0049 (1.29%) | 0.0045 (7.72%) | 0.0015 | 0.0015 (2.26%) | 0.0014 (7.66%) |
| | $t=15\text{ s}$ | 0.0036 | 0.0036 (0.44%) | 0.0034 (5.30%) | 0.0044 | 0.0044 (0.35%) | 0.0041 (5.42%) |
| | $t=20\text{ s}$ | 0.0025 | 0.0026 (3.84%) | 0.0025 (0.22%) | 0.0064 | 0.0066 (3.94%) | 0.0063 (0.19%) |
| CPU Times (s) | | 291375.17+45.98 | 1457.13+0.42 | 1507.92+0.44 | 291375.17+46.12 | 1457.13+0.39 | 1507.92+0.43 |

**Table 4**

Comparison of the computational results at point B of KCQ-WZ and KCQ-CL with the reference results.

| Index | Time | $w(B)$ | | | $\dot{w}(B)$ | | |
|---|---|---|---|---|---|---|---|
| | | KCQ-MC | KCQ-WZ | KCQ-CL | KCQ-MC | KCQ-WZ | KCQ-CL |
| KCQ-mean | $t=5\text{ s}$ | -0.0053 | -0.0053 (0.02%) | -0.0052 (0.42%) | -0.0047 | -0.0047 (0.03%) | -0.0047 (0.50%) |
| | $t=10\text{ s}$ | 0.0098 | 0.0098 (0.02%) | 0.0098 (0.58%) | -0.0037 | -0.0037 (0.03%) | -0.0037 (0.64%) |
| | $t=15\text{ s}$ | 0.0065 | 0.0065 (0.03%) | 0.0064 (0.51%) | -0.0078 | -0.0078 (0.03%) | -0.0077 (0.50%) |
| | $t=20\text{ s}$ | -0.0033 | -0.0033 (0.02%) | -0.0033 (0.34%) | -0.0085 | -0.0085 (0.02%) | -0.0084 (0.33%) |
| KCQ-SD | $t=5\text{ s}$ | 2.80E-04 | 2.78E-4 (0.86%) | 2.75E-4 (1.79%) | 3.83E-4 | 3.89E-4 (1.67%) | 3.79E-4 (0.93%) |
| | $t=10\text{ s}$ | 4.34E-04 | 4.14E-4 (4.50%) | 4.11E-4 (5.21%) | 1.96E-4 | 1.92E-4 (2.02%) | 1.87E-4 (4.40%) |

| | | | | | | |
|---|---|---|---|---|---|---|
| $t = 15$ s | 2.99E-04 | 2.91E-4 (2.42%) | 2.89E-4 (3.06%) | 3.66E-4 | 3.58E-4 (2.18%) | 3.55E-4 (3.10%) |
| $t = 20$ s | 1.88E-04 | 1.86E-4 (1.00%) | 1.90E-4 (1.03%) | 4.71E-4 | 4.67E-4 (0.92%) | 4.75E-4 (0.86%) |
| CPU Times (s) | 291375.17+47.86 | 1457.13+0.43 | 1507.92+0.41 | 291375.17+46.37 | 1457.13+0.40 | 1507.92+0.43 |

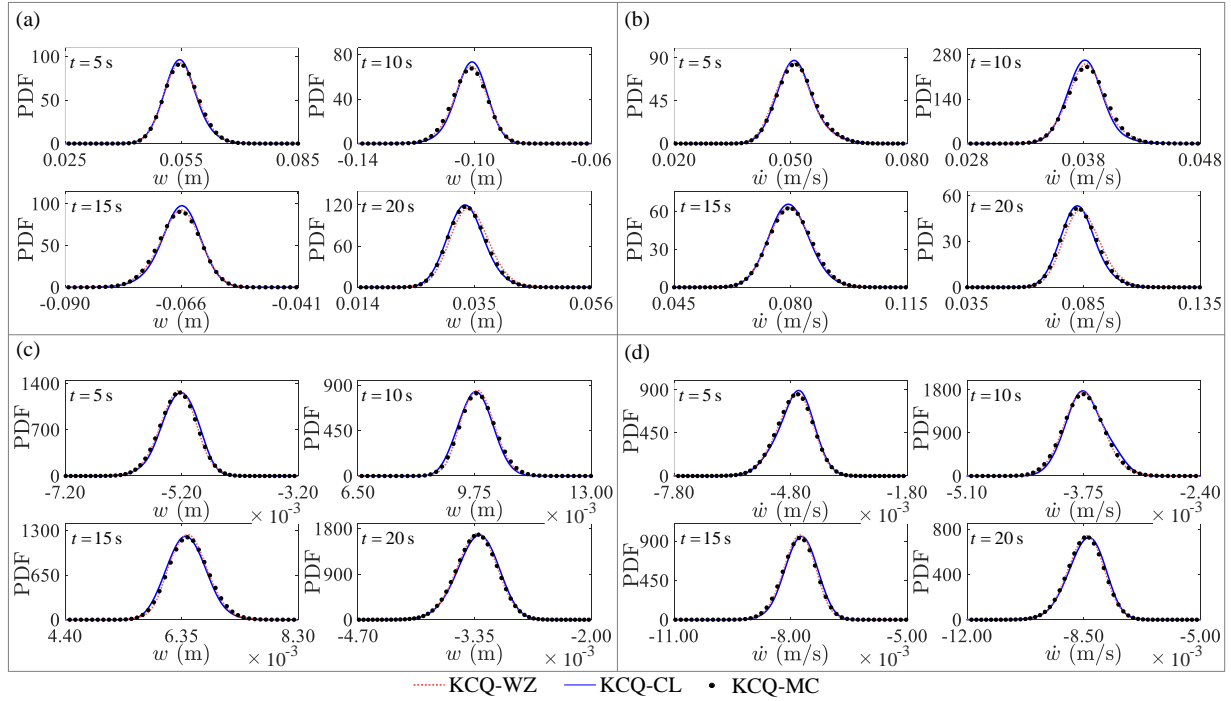

Fig. 17. Comparison of PDF curves of vertical displacement and velocity at points A and B at different times: (a) $w(A)$; (b) $\dot{w}(A)$; (c) $w(B)$; (d) $\dot{w}(B)$.

Table 3 and Table 4 respectively provide the random vertical displacement and vertical velocity at points A and B at time $t = 5, 10, 15, 20$ s calculated by KCQ-WZ and KCQ-CL, as well as the RE compared to the reference solution. The KCQ-PDF curves for the vertical displacement and vertical velocity at time $t = 5, 10, 15, 20$ s calculated using the KCQ-WZ, KCQ-CL, and KCQ-MC methods are plotted in Fig. 17.

It can be seen from Tables 3 and 4 that compared to the reference solution, the KCQ-mean and KCQ-SD calculated at each iteration step using KCQ-WZ both have very high accuracy, with REs all less than 5%. KCQ-CL occasionally has REs greater than 5% when calculating KCQ-SD, which is due to the small value of KCQ-SD itself (for example, the KCQ-SD for vertical displacement and deflection at point A reaches the order of 1E-3, and that for point B reaches the order of 1E-4). Although the RE is slightly greater than 5%, the absolute error obtained has reached the order of 1E-4 or 1E-5. At the same time, we have also recorded the CPU Times for KCQ-MC, KCQ-WZ, and KCQ-CL in Tables 3 and 4. The CPU Times results indicate that for the dynamic response of concrete bridges, both GQMC-based KCQ-WZ and KCQ-CL are significantly faster than the KCQ-MC method, with KCQ-WZ and KCQ-CL being approximately two orders of magnitude faster. Additionally, the offline database calculation steps for both KCQ methods account for about 99.99% of the entire computation process, thus the offline-online coupled computational strategy proposed in Section 4 is indeed effective in reducing the computational burden of the online calculation steps. From Fig. 17, it can be seen that the PDF curves for displacement and velocity at each iteration step calculated based on KCQ-WZ and KCQ-CL match well with the displacement and velocity PDF curves obtained from KCQ-MC calculations. In summary, the proposed KCQ method can accurately estimate the uncertainty of the random response of the concrete bridges considering measurement conditions.

Next, the traditional N-MC method, which does not consider measurement conditions, is used to calculate the mean, SD, and PDF of the displacement and velocity at points A and B, and these results are compared with the KCQ-mean, KCQ-SD, and KCQ-PDF obtained from the KCQ-WZ calculations. First, the TSDI for the vertical displacement and velocity at points A and B calculated by KCQ-WZ and N-MC is compared, and the results are plotted in Fig. 18. It is very clear from Fig. 18 that when calculating the vertical displacement $w$ and velocity $\dot{w}$, the TSDI obtained by the KCQ-WZ method, which includes measurement conditions, is still narrower than the TSDI obtained by the N-MC method. We further compared the PDFs of the vertical displacement and velocity at time $t = 5, 10, 15, 20 \text{ s}$ calculated by the KCQ-WZ and N-MC methods, and the results are plotted in Fig. 19. It is evident that the PDF curves obtained by the KCQ-WZ method, which considers measurement conditions, are significantly narrower than those obtained by the traditional N-MC method, which does not consider measurement conditions. In summary, for the considered random vibration problem of the concrete bridge, the responses calculated by the KCQ method, which takes into account measurement conditions, are more tightly clustered around their mean and have a smaller SD.

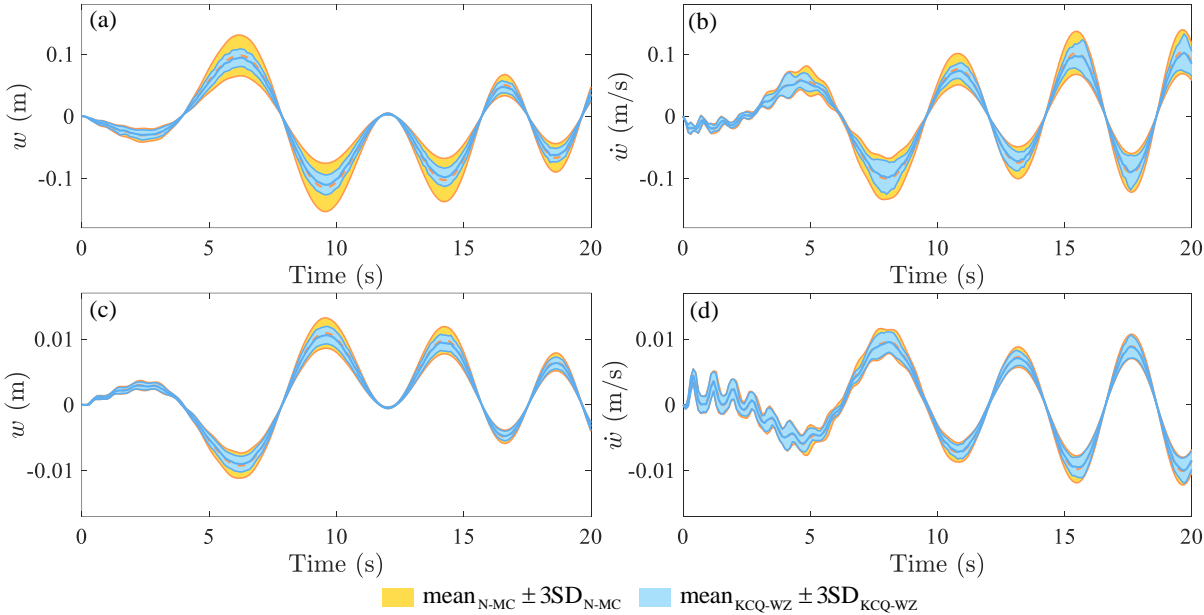

Fig. 18. TSDIs of KCQ-WZ and N-MC for vertical displacement and velocity at points A and B: (a) $w(A)$; (b) $\dot{w}(A)$; (c) $w(B)$; (d) $\dot{w}(B)$.

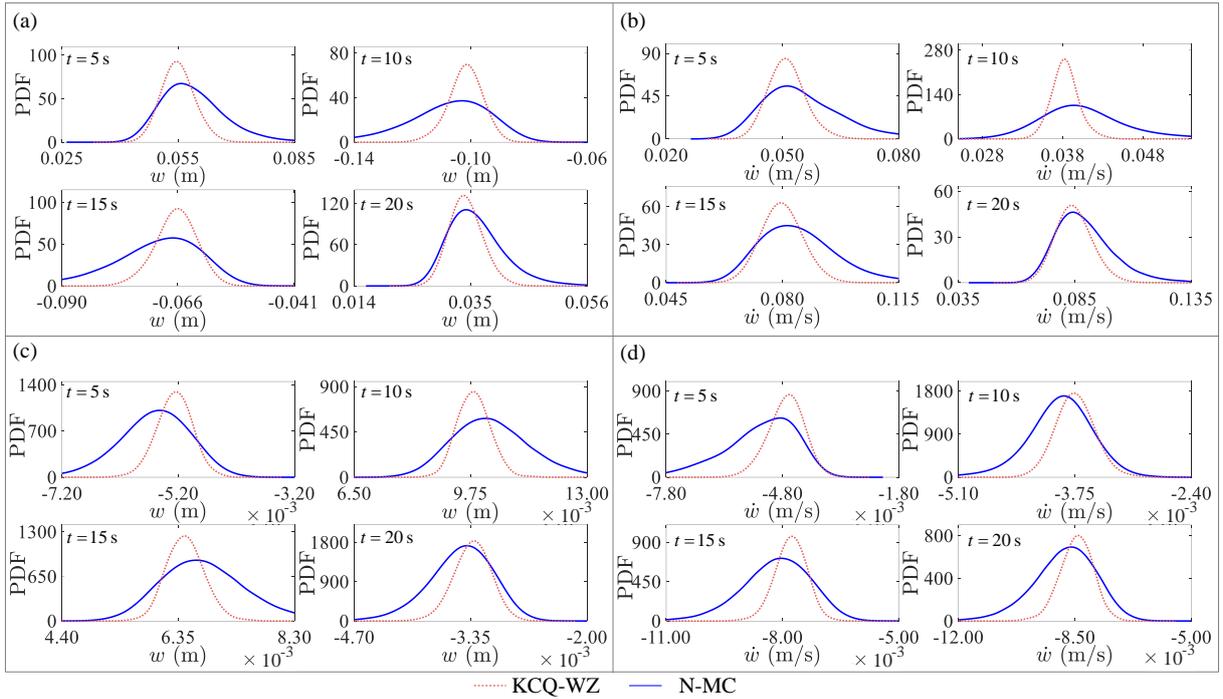

Fig. 19. PDF curves for vertical displacement and velocity at points A and B at various times, calculated with and without considering measurement conditions: (a) $w(\mathrm{A})$; (b) $\dot{w}(\mathrm{A})$; (c) $w(\mathrm{B})$; (d) $\dot{w}(\mathrm{B})$.

## 6. Conclusions

This paper proposes a new theory of key condition quotient (KCQ) for the uncertain quantification of random structural dynamic responses under the condition of measurement data containing random errors. The theory extracts key measurement conditions from the measurement data by measuring the correlation between the measurement data and the random response. Based on the principle of probability conservation and conditional probability theory, it provides exact expressions in the form of quotients for the conditional probability density function (PDF), conditional mean, and conditional variance of the structural random response under the condition of key measurement data, respectively referred to as KCQ-PDF, KCQ-mean, and KCQ-variance. The paper also presents a numerical format for calculating these three KCQs by combining the generalized polynomial chaos method (GQMC) with the smoothing of Dirac functions. Furthermore, an efficient numerical algorithm for the uncertain quantification of structural random response under the condition of key measurement data is constructed through an offline-online coupled computational strategy.

The established key condition quotient (KCQ) numerical algorithm is validated through three numerical examples: a single-degree-of-freedom spring-mass-damper system, a geometrically nonlinear beam, and a concrete cable-stayed bridge. During the numerical verification process, the GQMC format proposed by Chen and Li, and the GQMC format proposed by Wu and Zhao were used to construct the numerical calculation format for KCQ (referred to as KCQ-CL and KCQ-WZ in this paper, respectively). The Monte Carlo method is also employed to calculate the high-dimensional numerical integration in KCQ, denoted as KCQ-MC, and the result of this method is used as the reference solution. Numerical comparison results show that KCQ-WZ and KCQ-CL, which use the GQMC method, can achieve computational efficiency much faster than KCQ-MC while meeting the requirements for computational accuracy, enabling the efficient and precise assessment of the uncertainty in structural random dynamic responses considering measurement conditions. The examples also compare the traditional non-conditional uncertain quantification results with the conditional uncertain quantification results that take into account key measurement information, demonstrating that the KCQ algorithm proposed in this paper, which considers measurement conditions, can significantly reduce the uncertainty of the estimated dynamic responses, providing a more refined statistical basis for structural safety and reliability analysis.

Although the numerical examples demonstrate that the proposed KCQ algorithm performs well in conditional uncertain quantification analysis of dynamic responses affected by uncertain loads and structural parameters, the current KCQ algorithm does not yet consider issues such as non-independent measurement errors and unknown distribution of uncertain loads and structural parameters. Considering the importance of the aforementioned issues in real-world engineering, further research will be conducted. We believe that the proposed KCQ algorithm will play a significant role in the refined quantification of dynamic responses involving uncertainty in the future.

## Appendix (The principle of probability conservation)

The principle of probability conservation has been studied in-depth by Li and Chen et al. in Refs. [40]. It says in the stochastic dynamical system, the principle of probability conservation can be described by the following generalized probability density evolution equation:

$$\frac{\partial \rho_{\alpha,u}(\alpha,u,t)}{\partial t} + \dot{\varphi}(\alpha,t) \frac{\partial \rho_{\alpha,u}(\alpha,u,t)}{\partial u} = 0 \tag{A1}$$

where $u = \varphi(\alpha,t)$ is the random dynamical response, $\alpha$ is the random parameter, and $\rho_{\alpha,u}$ is the joint PDF. Integrating Eq. (A1) yields the following equation

$$\rho_{\alpha,u}(\alpha,u,t) = \rho_\alpha(\alpha) \delta(u - \varphi(\alpha,t)) \tag{A2}$$

where $\rho_\alpha(\alpha)$ is the PDF of $\alpha$. Equations (A1) and (A2) are equivalent to each other, and both equations can describe the probability conservation in the stochastic dynamical system. Here, we will extend the principle of probability conservation to the general stochastic problem, as shown in the following theorem.

**Theorem 1** *(Principle of Probability Conservation). There exist two any random vectors $\zeta$ and $\alpha$, where the PDF of $\alpha$ is $\rho_\alpha(\alpha)$, and $\zeta$ can be expressed in terms of $\alpha$ as follows:*

$$\zeta = (\zeta_1, \zeta_2, \cdots, \zeta_s)^T = [f_1(\alpha), f_2(\alpha), \cdots, f_s(\alpha)]^T = f(\alpha), \tag{A3}$$

*then the joint PDF of $\zeta$ and $\alpha$ is:*

$$\rho_{\zeta,\alpha}(\zeta,\alpha) = \rho_\alpha(\alpha) \delta(\zeta - f(\alpha)), \tag{A4}$$

where $\delta(\boldsymbol{\zeta}-\boldsymbol{f}(\boldsymbol{\alpha}))=\prod_{i=1}^{s}\delta(\zeta_i-f_i(\boldsymbol{\alpha}))$.

**Proof.** Without loss of generality, let $\boldsymbol{a}=(a_1,a_2,\cdots,a_s)^{\mathrm{T}}$ be an any random vector, then we have

$$P(\boldsymbol{\zeta}\leq\boldsymbol{a})=\int_{-\infty}^{a}\left[\int_{-\infty}^{+\infty}\rho_{\zeta,\alpha}(\boldsymbol{\zeta},\boldsymbol{\alpha})\mathrm{d}\boldsymbol{\alpha}\right]\mathrm{d}\boldsymbol{\zeta}=\int_{-\infty}^{+\infty}\int_{-\infty}^{a}\rho_{\zeta,\alpha}(\boldsymbol{\zeta},\boldsymbol{\alpha})\mathrm{d}\boldsymbol{\zeta}\mathrm{d}\boldsymbol{\alpha}, \quad (A5)$$

where $P(\boldsymbol{\zeta}\leq\boldsymbol{a})$ represents the probability that $\boldsymbol{\zeta}\leq\boldsymbol{a}$.

Note that $\boldsymbol{f}$ is a function maps the random variable $\boldsymbol{\alpha}$ to the random variable $\boldsymbol{\zeta}$. According to the principle of probability conservation, the probability of $\boldsymbol{\zeta}\leq\boldsymbol{a}$ is essentially the probability of $\boldsymbol{f}(\boldsymbol{\alpha})\leq\boldsymbol{a}$, so there is:

$$P(\boldsymbol{\zeta}\leq\boldsymbol{a})=P(\boldsymbol{f}(\boldsymbol{\alpha})\leq\boldsymbol{a})=\int_{-\infty}^{+\infty}I(\boldsymbol{f}(\boldsymbol{\alpha})\leq\boldsymbol{a})\rho_\alpha(\boldsymbol{\alpha})\mathrm{d}\boldsymbol{\alpha}, \quad (A6)$$

where, $I(\bullet)$ is the characteristic function [59]. By combining Eqs. (A5) and (A6), we can obtain:

$$\int_{-\infty}^{a}\rho_{\zeta,\alpha}(\boldsymbol{\zeta},\boldsymbol{\alpha})\mathrm{d}\boldsymbol{\zeta}=I(\boldsymbol{f}(\boldsymbol{\alpha})\leq\boldsymbol{a})\rho_\alpha(\boldsymbol{\alpha}). \quad (A7)$$

Because $\boldsymbol{a}-\boldsymbol{f}(\boldsymbol{\alpha})\in[\boldsymbol{0},\infty)$,

$$I(\boldsymbol{f}(\boldsymbol{\alpha})\leq\boldsymbol{a})=\int_{-\infty}^{a_1}\int_{-\infty}^{a_2}\cdots\int_{-\infty}^{a_n}\delta(\boldsymbol{\zeta}-\boldsymbol{f}(\boldsymbol{\alpha}))\mathrm{d}\boldsymbol{\zeta}.$$

Taking the derivative of $\boldsymbol{a}$ on both sides of Eq. (A7), there is

$$\rho_{a,\alpha}(\boldsymbol{a},\boldsymbol{\alpha})=\rho_\alpha(\boldsymbol{\alpha})\delta(\boldsymbol{a}-\boldsymbol{f}(\boldsymbol{\alpha})),$$

at this point, $\boldsymbol{a}$ can be replaced by any vector, so letting $\boldsymbol{a}=\boldsymbol{\zeta}$ yields Eq. (A4). ∎

## CRediT authorship contribution statement

Feng Wu: Writing – original draft, Conceptualization, Validation, Software, Methodology,

Supervision, Project administration, Funding acquisition, Resources. Yue-Lin Zhao: Writing – original draft, Conceptualization, Visualization, Methodology, Data curation, Software, Validation, Formal analysis, Project administration. Li Zhu: Writing – original draft, Visualization, Data curation, Software, Validation, Formal analysis.

## Declaration of competing interest

The authors declare that they have no known competing financial interests or personal relationships that could have appeared to influence the work reported in this paper.

## Data availability

Data will be made available on request.

## Acknowledgments

This work was supported by the National Natural Science Foundation of China (Nos. 12372190, 62388101), Fundamental Research Funds for the Central Universities (Nos. DUT20RC (5)009, DUT20GJ216), Natural Science Foundation of Liaoning Province (No. 2021-MS-119).